\documentclass{aa}
\usepackage{graphics}
\usepackage[latin1]{inputenc}

\newcommand{\kms}   {~km~s$^{-1}$}
\newcommand{\mjy}   {~mJy~beam$^{-1}$}

\newcommand{\cmd}   {~cm$^{-2}$}
\newcommand{\cmt}   {~cm$^{-3}$}
\newcommand{\vlsr}  {$v_{\rm LSR}$}
\newcommand{\lo}    {$L_{\sun}$}
\newcommand{\mo}    {$M_{\sun}$}
\newcommand{\T}[1]  {T_{\rm #1}}
\newcommand{\et}    {et al.}
\newcommand{\eg}    {e.\,g.,}
\newcommand{\ie}    {i.\,e.,}
\newcommand{\hhd}   {HH~2}
\newcommand{\nh}    {NH$_3$}
\newcommand{\h}     {H$_2$}
\newcommand{\hco}   {HCO$^+$}
\newcommand{\htco}  {H$^{13}$CO$^+$}
\newcommand{\htcn}  {H$^{13}$CN}
\newcommand{\form}  {H$_2$CO}
\newcommand{\metha} {CH$_3$OH}
\newcommand{\chtcn} {CH$_3$CN}
\newcommand{\cthd}  {C$_3$H$_2$}
\newcommand{\cdo}   {C$^{18}$O}
\newcommand{\tco}   {$^{13}$CO}
\newcommand{\J}[2]  {\mbox{#1--#2}}
\newcommand{\JK}[4] {\mbox{#1$_{#2}$--#3$_{#4}$}}
\newcommand{\Jp}[4] {\mbox{$\frac{#1}{#2},\frac{#3}{#4}$}}
\newcommand{\BM}[3] {\mbox{#1$\times$#2,~$#3$}}
\newcommand{\ap}    {A$^{+}$}
\newcommand{\mq}    {$\la$}
\newcommand{\N}[1]  {$\times10^{#1}$}

\newcommand{\arcdeg}{\mbox{$^\circ$}} 
\newcommand{\nodata}{ ~$\cdots$~ }    

\begin{document}


\title{The Molecular Condensations Ahead of Herbig-Haro Objects. III. 
Radiative and dynamical perturbations of the HH 2 condensation.}

\author{
Josep Miquel Girart\inst{1,2,3}, 
Serena Viti\inst{4},
Robert  Estalella\inst{3}, 
David A. Williams\inst{4}
}

\offprints{J.M. Girart: girart@ieec.fcr.es}

\institute{
Institut de Ci\`encies de l'Espai (CSIC),
Gran Capit\`a 2, 08034 Barcelona, Catalunya, Spain
\and
Institut d'Estudis Espacials de Catalunya
\and
Departament d'Astronomia i Meteorologia, Universitat de Barcelona,  
Av.\ Diagonal 647, 08028 Barcelona, Catalunya, Spain
\and
Department of Physics and Astronomy, University College London,
London, WC1E 6BT, England, UK
}

\date{Received ...; accepted ...}

\abstract{

We have carried out an extensive observational study (from BIMA data) and made
a preliminary theoretical investigation of the molecular gas around \hhd. The
molecular maps show a very complex morphological, kinematical and chemical
structure. For clarity we divided the observed region in four subregions: (1)
The Ahead Core, located ahead of \hhd: its chemistry may be a consequence of a
weak UV field originating in \hhd.  The chemical structure within the Ahead
Core suggests that it is not homogeneous but probably composed of small clumps;
(2) The SO$_2$ Clump, which is a molecular component within the Ahead Core that
is more exposed to the UV radiation from \hhd.  An increase of density and
relative molecular abundances is observed towards \hhd. The UV radiation is
possibly the source of molecular enhancement.  Our chemical analysis confirms
that this clump must have substructure within it; (3) the West Core, which is
surrounded by a ring structure of shocked ionized gas and mid--IR emission. The
ring structure is likely a consequence of the fact that the core is in the
foreground with respect to the shocked and hot component.  The chemistry of
this core can be best explained as arising from a combination of an old
photo--processed dense clump and a PDR, with or without a warm interface
created in the interaction of the outflow with the core; (4) The High Velocity
Region, associated with \hhd, is traced by \hco\ but not by other molecular
shock tracers.  The chemistry can be accounted for by the interaction of the
VLA~1 outflow with a dense clump $ via$ non--dissociative shocks and by the
presence of a very strong UV field. The overall main conclusion of this work
confirms the findings of Paper I and II, by demonstrating that in addition to
the strong photochemical effects caused by penetration of the UV photons from
HH~2  into molecular cloud, a range of complex radiative and dynamical
interactions occur.  Thus, despite the apparent `quiescent' nature of the
molecular cloud ahead of \hhd, the kinematical properties observed within the
field of view suggest that it is possibly being driven out by the powerful
winds from the VLA~1 protostar.

}

\authorrunning{Girart et al.}
\titlerunning{Radiative and dynamical perturbations around HH 2}

\maketitle


\section{Introduction\label{intro}}

Very powerful winds are associated with the earliest stages of star formation. 
These winds alter significantly the dense molecular environments that surround
the protostars. There are two kinds of interaction between the winds with the
dense molecular environment; one is dynamic, and the other radiative. The
molecular outflows are the main signpost of the dynamical interaction (\eg\
Richer \et\ \cite{Richer00}). The radiation generated in the strong shocks
produced in the Herbig--Haro (HH) objects, as well as the shock themselves
alter the chemical composition of the molecular gas through mantle desorption
and gas phase endothermic reactions (\eg\ Girart \et\ \cite{Girart94},
\cite{Girart02};  Flower \et\ \cite{Flower96}; Amin \cite{Amin01}). Thus, it is
clear that high angular and spectral resolution maps of the molecular gas where
this interaction takes place are very important in order to understand better 
the interaction mechanism and the properties of both the outflow and of the
molecular clouds.

%
     \begin{table*}
     \caption[]{Frequency setups of the BIMA observations}
     \label{tobs}
     \[
     \begin{tabular}{c@{\hspace{0cm}}l@{\hspace{0.3cm}}l@{\hspace{0cm}}c}
     \noalign{\smallskip}
     \hline
     \noalign{\smallskip}   
\multicolumn{1}{c}{Frequency (GHz)} &
\multicolumn{2}{c}{} & 
\multicolumn{1}{c}{System} 
\\
\multicolumn{1}{c}{LSB -- USB} &
\multicolumn{1}{l}{Main molecular  transitions observed} &
\multicolumn{1}{l}{Dates} &
\multicolumn{1}{c}{Temperatures}
\\
     \noalign{\smallskip}
     \hline
     \noalign{\smallskip}     
     $\!$72.2~-- 75.8
& DCO$^+$ 1--0, SO$_2$ \JK{6}{0,6}{5}{1,5}, DCN 1--0, CH$_3$SH \JK{3}{0}{2}{0}\ A$^+$
& 2000 Nov, 2001 Mar 
& 250--1000~K \\
     $\!$72.7~-- 76.4
& \form\ \JK{1}{0,1}{0}{0,0}, SO$_2$ \JK{6}{0,6}{5}{1,5}, DCN 1--0
& 2000 Apr          
& 300--~850~K \\
     $\!$85.3~-- 89.0
& HCN 1--0, \hco\ 1--0, \htco\ 1--0, \cthd\ \JK{2}{1,2}{1}{0,1}, HCS$^+$ 2--1
& 2000 Apr         
& 200--~600~K \\
     $\!$86.1~-- 90.0
& SO \JK{2}{2}{1}{1}, HCOOH \JK{4}{0,4}{3}{0,3}
& 2000 Jul 
& 200--~500~K \\
     $\!$86.7~-- 89.7
& \htco\ \J{1}{0}, \htcn\ \J{1}{0}, SiO 2--1
& 2003 May, Jun$^a$
& 170--~600~K \\
96.7~--100.0
& \metha\ \JK{2}{n}{1}{n}, HC$_3$N 11--10
& 1999 Oct          
& 200--~500~K \\
98.0~--101.0
& CS \J{2}{1}, OCS 8--7
& 1999 Dec, 2000 May 
& 200--~800~K \\
99.3~--103.0
& SO \JK{3}{2}{2}{1}, H$_2$CS \JK{3}{0,3}{2}{0,2}
& 2000 Jul 
& 300--1400~K \\
$\!\!\!$100.5~--104.1
& SO$_2$ \JK{3}{1,3}{2}{0,2}, NH$_2$CN \JK{5}{1,4}{4}{1,3}
& 2000 Nov, 2001 Mar, May
& 140--~400~K \\
$\!\!\!$104.2~--107.8
& SO$_2$ \JK{3}{1,3}{2}{0,2} 
& 2000 Jul 
& 300--~800~K \\
$\!\!\!$109.7~--113.0
& SO \JK{2}{3}{1}{2}, \cdo\ 1--0, CN 1--0
& 2000 Apr, May 
& 280--1300~K \\
$\!\!\!$109.9~--113.4
& \cdo\ 1--0, \tco\ 1--0, CN 1--0, \chtcn\ \JK{6}{n}{5}{n}, HNCO \JK{5}{0,5}{4}{0,4}
& 2000 Nov, 2001 Mar--May 
& 350--1100~K \\
     \noalign{\smallskip}
     \hline
     \end{tabular}
     \]
     \begin{list}{}{}
\item[$^a$] 2003 June observations were carried out in the D configuration
     \end{list}
    \end{table*}
%

\hhd\ is a well studied bright, high excitation HH object, which generates
strong UV radiation  (\eg\ B\"ohm \et\ \cite{Boehm92};  Raymond \et\
\cite{Raymond97}) and presents a very complex morphology (Hester \et\
\cite{Hester98};   Bally \et\ \cite{Bally02}).   It is driven by  HH~1--2 VLA~1
(hereafter VLA~1), a Class 0 protostar  (Andr\'e, Ward--Thompson \& Barsony
\cite{Andre00}). \hhd\ is also associated with CO emission from the  VLA~1
molecular outflow (Moro--Mart{\' i}n \et\ \cite{Moro99}).    A dense cloud of
molecular gas and dust, apparently quiescent, appears  ahead of  \hhd\ (Davis,
Dent \& Bell Burnell \cite{Davis90}; Torrelles \et\ \cite{Chema92}; Dent,
Furuya \& Davis \cite{Dent03}).

Girart \et\ (\cite{Girart02}; hereafter Paper I) carried out a molecular line
survey in the 0.8--4.0 mm range and at an angular resolution of $\sim 30''$ of
the dense molecular gas ahead of \hhd. From these observations we derive a
temperature of $\sim 13$~K and a density of 3\N{5}~\cmt. The observations show
a characteristic chemistry that can be accounted for with chemical models of
irradiated clumps (\eg\ Taylor \& Williams \cite{Taylor96}; Viti \& Williams
\cite{Viti99});  the models indicate that photochemistry induced by HH
radiation in regions that have not yet been shocked should produce a
characteristic chemistry that is a signature of that origin.  A detailed
modeling of the conditions of the molecular clump ahead of HH~2 was carried out
by  Viti \et\ (\cite{Viti03}; hereafter Paper II). From the chemical analysis
of this work we found that the  `illuminated' clump was young when it was first
irradiated by \hhd\ and that to account for the observed column densities a
density gradient is required (\ie\ the clump does not have a uniform density).
In addition, we found that the peculiar chemistry will last for no more than a
few hundred years, since the UV radiation will ultimately destroy the molecular
content.  The chemical effects, however, last longer as the relevant region
`eats' into the core.

In Paper I we studied the observational properties and the molecular gas at an
angular resolution of 30$''$ at one position ahead of \hhd, by using data from
the CSO and BIMA. In Paper II we carried out chemical analysis by using the
results of Paper I. The aim of this work is to study in detail, at higher
angular resolution ($\sim 10''$), the distribution and properties of the
molecular gas around \hhd\ and to understand how the powerful outflow arising
from VLA~1 affects its properties. In \S~2 we summarize the observational
parameters of the maps. In \S~3 we describe the morphological properties of
the  molecular emission.   In \S~4 we analyze the data. 
In \S~5 we discuss the
chemical as well as the physical properties of the molecular environment around
\hhd\ and we consider why the regions observed appear to be chemically
distinct.  A brief summary  and the conclusions are given in \S~6.

%
     \begin{table}
     \caption[]{Parameters of the BIMA maps}
     \label{tbima}
     \[
\begin{tabular}{l@{\hspace{0cm}}c@{\hspace{0cm}}r@{\hspace{0.1cm}}c@{\hspace{0cm}}c@{\hspace{0cm}}}
     \noalign{\smallskip}
     \hline
     \noalign{\smallskip}   
\multicolumn{2}{c}{} &
\multicolumn{1}{c}{Synthesized} & 
\multicolumn{2}{c}{} 
\\
\multicolumn{2}{c}{} &
\multicolumn{1}{c}{Beam} & 
\multicolumn{1}{c}{} &
\multicolumn{1}{c}{$rms$} 
\\
\multicolumn{1}{c}{Molecule} &
\multicolumn{1}{c}{$\nu$} &
\multicolumn{1}{c}{HPBW, PA} &
\multicolumn{1}{c}{$\Delta v$} & 
\multicolumn{1}{c}{(Jy/} 
\\
\multicolumn{1}{c}{Transition} &
\multicolumn{1}{c}{(GHz)} &
\multicolumn{1}{c}{$\!\!$(arcsec, deg)} &
\multicolumn{1}{c}{$\!\!$(km s$^{-1}$)} & 
\multicolumn{1}{c}{$\!\!$beam)} 
\\
     \noalign{\smallskip}
     \hline
     \noalign{\smallskip}     
DCO$^+$ \J{1}{0}    & 72.0393& \BM{19.3}{7.3}{1} &0.20&0.30 \\
H$_2$CO \JK{1}{0,1}{0}{0,0}& 72.8380& \BM{18.0}{9.4}{15}&0.20&0.27 \\
\cthd\  \JK{2}{1,2}{1}{0,1} & 85.3389& \BM{14.1}{7.1}{-1}&0.34&0.10 \\
SO      \JK{2}{2}{1}{1}    & 86.0934& \BM{12.9}{7.6}{0} &0.34&0.16 \\
\htcn\  \J{1}{0}    & 86.3402& \BM{17.6}{11.8}{4}&0.33&0.11 \\
\htco\  \J{1}{0}    & 86.7543& \BM{17.4}{11.2}{9}&0.33&0.11 \\
SiO     \J{2}{1}    & 86.8470& \BM{17.7}{11.7}{6}&0.33&0.11 \\
HCN     \J{1}{0}    & 88.6318$^a$& \BM{12.7}{7.1}{0} &0.33&0.10 \\
\hco\   \J{1}{0}    & 89.1885& \BM{13.2}{7.0}{1} &0.33&0.11 \\
\metha\ \JK{2}{n}{1}{n}& 96.7414$^b$& \BM{11.7}{6.4}{4} &0.30&0.13 \\
CS      \J{2}{1}    & 97.9810& \BM{14.2}{6.1}{10}&0.30&0.14 \\
SO      \JK{3}{2}{2}{1}    & 99.2999& \BM{11.9}{6.6}{15}&0.59&0.14 \\
SO$_2$  \JK{3}{1,3}{2}{0,2}&104.0294&  \BM{9.9}{5.6}{1} &0.28&0.05 \\
SO      \JK{2}{3}{1}{2}    &109.2522& \BM{10.4}{5.9}{2} &0.27&0.15 \\
\cdo\   \J{1}{0}    &109.7822& \BM{10.1}{5.9}{1} &0.27&0.12 \\
\tco\   \J{1}{0}    &110.2014&  \BM{9.7}{6.3}{-8}&0.27&0.13 \\
CN      \J{1}{0}    &113.4910$^c$& \BM{12.0}{8.9}{-5}&0.52&0.08 \\
     \noalign{\smallskip}
     \hline
     \end{tabular}
     \]
     \begin{list}{}{}
\item[$^{a}$] Frequency for the F=2--1 hyperfine line. Observations also 
detected the F=1--1 and F=0--1 hyperfine lines
\item[$^{b}$] Frequency for the \JK{2}{0}{1}{0}\ap\ line. Observations 
also detected the \JK{2}{-1}{1}{-1} E line
\item[$^{c}$] Frequency for the \J{1}{0} \Jp{3}{2}{5}{2}--\Jp{1}{2}{3}{2} line.
Observations also detected the \Jp{3}{2}{1}{2}--\Jp{1}{2}{1}{2} and
\Jp{1}{2}{3}{2}--\Jp{1}{2}{3}{2} lines. The maps were obtained by applying a
Gaussian taper of 26~k$\lambda$ to the visibilities.
     \end{list}
    \end{table}
%

%
\setcounter{figure}{3}
\begin{figure} 
\begin{center}
\resizebox{7.5cm}{!}{\includegraphics{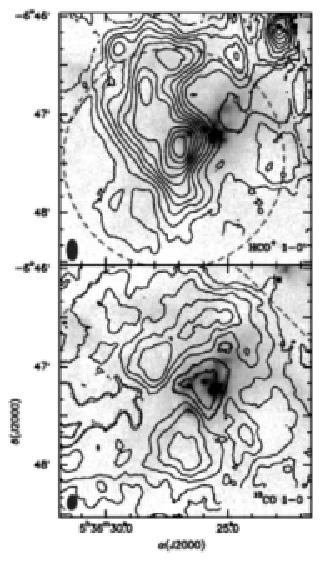}}
\end{center}
     \hfill
     \caption[]{     
     Superposition of the [SII] image (from Curiel, private communication) and
the BIMA contour maps of the zero--order moment (integrated emission  over the 
5.3 to 11.3~\kms\ velocity range)  of the \hco\ \J{1}{0}\ and \tco\ \J{1}{0}\
around HH~2.  The BIMA maps are  corrected by the primary beam response.
Contours are 1, 2.5, 4, \dots, 32.5 times  0.45 (\hco) and 0.72 (\tco) Jy~\kms.
The contours with dots around \hhd\ indicates an  emission valley.  The dashed
thick line shows the BIMA beam response at the 0.5 level with respect to phase
center.  The synthesized beam is shown in the  bottom left corner of the
panels.  The position of the driving source of the  HH~1--2 outflow (VLA~1) is
shown as a filled square in the upper--right corner  of the panels.
}
     \label{fmom} 
\end{figure}

\section{Observations}

The observations were carried out with the 10--antenna BIMA array\footnote{The
BIMA array is operated by the Berkeley--Illinois--Maryland Association  with
support from the National Science Foundation.} at the Hat Creek Radio
Observatory.  The phase calibrators used were QSOs 0541$-$056, 0607$-$085 and
0609$-$157.   Absolute flux calibration was checked by observing Saturn or
Mars.   A total of 12 frequency setups were used, which covered frequencies
between 72.0 and 113.5~GHz. All the frequency setups were observed in the C
configuration  (\ie\ antenna separation from 6.3 to 100~m).  At 3~mm this
configuration provides a visibility coverage between 20 and 300~nanoseconds,
\ie\ an angular resolution of $\sim 9''$.  Table~\ref{tobs} shows for each
frequency setup, the frequencies of the lower and upper side band (LSB and USB),
the main molecular transitions observed, the observation dates and the range of
system temperatures achieved. The digital correlator was configured to sample
part of the 800~MHz wide IF passband in several windows, with adjustable
frequency resolution.  The typical window for the line observations was
configured with a 25~MHz bandwidth and 256 channels, giving a spectral 
resolution of 97.7~kHz.   Thus, with the powerful correlator we were able to
observe many molecular transitions.  Table~2 from Paper I gives the list of the
lines observed before 2003, with the peak intensity or upper limits for an
angular resolution of 30$''$. In the 2003 observations, of the three lines
observed (see Table~\ref{tobs}), we only detected the \htco\ \J{1}{0}.  

The phase center of the observations was  $\alpha (J2000) = 5^{\rm h}36^{\rm
m}27\fs20$;  $\delta (J2000) = -6\arcdeg47'27''$. The FWHM BIMA primary beam
ranges from $2\farcm7$ at 72~GHz to $1\farcm7$ at  113~GHz, and thus our
observations engulfs within the primary beam the \hhd\ object and dense core
associated with it.  For the 109.9--113.4~GHz (LSB--USB)  frequency setup, the
observations (which included the \tco, \cdo\ and the CN lines) were done with a
mosaic of five points: a field centered in the phase center, and the other
fields located 48$''$ N, S, E and W of the the phase center. The data were
calibrated using the MIRIAD software.  Maps were made with the $(u,v)$ data
weighted by the associated system temperatures and using natural weighting.
Table~\ref{tbima} lists all the molecular lines detected from the observations
listed in Paper I as well as the three lines observed in 2003. The table
includes the channel resolution, the resulting synthesized beam and the $rms$
noise of the maps  at this channel resolution.

\section{Results}

Figures~\ref{fcanalA}, \ref{fcanalB} and \ref{fcanalC} show the channel maps 
over the 6 to 10~\kms\ velocity range of the molecular transitions with
emission detected at least in one channel. For most of the molecules detected
their emission is comprised within this velocity range. Figure~\ref{fmom}
shows  the zero--order moment of the emission for the \tco\ and \hco\
\J{1}{0}\  lines. The kinematical and morphological complexity of this region
is clear from these figures. 

\begin{figure} 
\begin{center}
\resizebox{8.0cm}{!}{\includegraphics{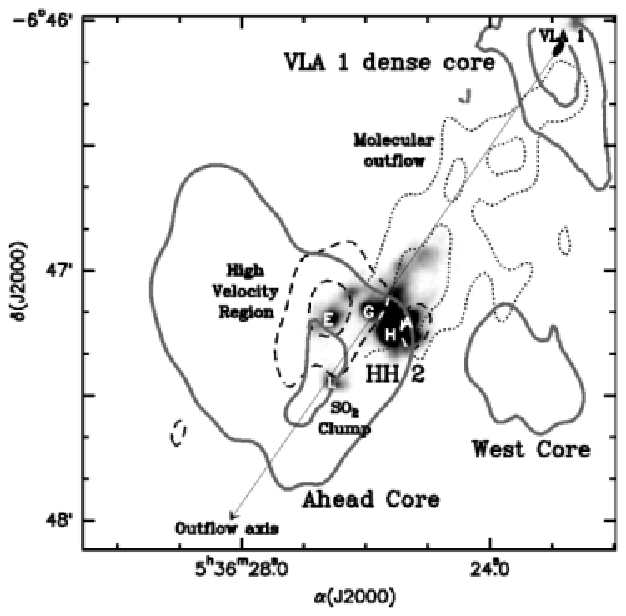}}
\end{center}
     \hfill
     \caption[]{     
     Sketch of the \hhd\ region. Thick grey contours show the main molecular
regions (the Ahead and West Cores with respect to \hhd\ and the VLA~1 dense 
core). The dashed contours show the high velocity \hco\ clump. The dotted
points contours show the jet--like molecular  outflow (obtained from the \hco\
emission at a \vlsr\ velocity of 7.3~\kms\  after masking the contribution from
the Ahead Core).  The  grey scale shows the [SII] emission (from Curiel,
private communication). The white labels within the [SII] emission indicates
the \hhd\ knot names.  The small filled ellipsoid  shows the position and
elongation direction of  the VLA~1 radio jet, the  powering source of the
HH~1--2 outflow.   The arrow shows the direction of the outflow.
}
     \label{fesquema} 
\end{figure}
%

For clarity, we will divide the observed region in four subregions taking into
account the properties of the different molecular species detected and the
properties of other tracers detected at other wavelengths (see
Figure~\ref{fesquema}): 
(1) The molecular core ahead of \hhd\ (hereafter Ahead Core); this region
includes  the region  studied in Paper I and II, which coincides with the
compact emission traced by SO and SO$_2$; 
(2) The molecular core west of \hhd\ (hereafter West Core); 
(3) The high velocity  molecular emission spatially associated with \hhd\
(hereafter  High Velocity Region); 
(4) The dense core surrounding the HH~2 driving source, VLA~1.  In the following
three subsections we describe the first three regions. The VLA~1 dense core is
detected clearly in several molecules (\tco, \htco, \form,  HCN, \cthd) in the
10~\kms\ channel map (see Figs.~\ref{fcanalA},  \ref{fcanalB} and
\ref{fcanalC}).  However, since it is  located far from the  phase center of the
BIMA observations ($\sim 1\farcm7$) we do not study its  properties. The
properties of this core have been well described in previous studies  (\eg\
Torrelles \et\ \cite{Chema94}; Choi \& Zhou \cite{Choi97}).  Thus, below we
describe the first three regions. We also describe the overall kinematics of the
dense gas seen within the field of view. In the last subsection of the results
we briefly describe the emission traced by the \tco\ \J{1}{0}.

%
     \begin{figure*} 
    \resizebox{\hsize}{!}{\includegraphics{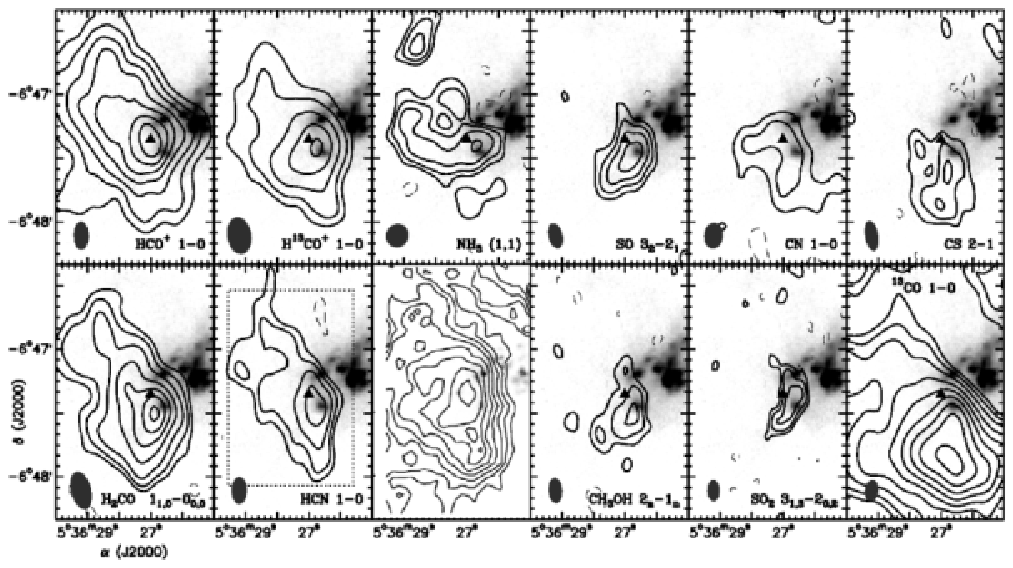}}
     \caption[]{
Contour maps of the ahead core for the dust emission at 0.85~mm (adapted from
Figure~2 of Dent \et\ \cite{Dent03}) and for several molecular lines. The  grey
scale [SII] image (from Curiel, private communication) is also shown in the
molecular line panels.  The molecular maps shown here are obtained by averaging
the emission over the 5.2 to 7.8~\kms\ \vlsr\ velocity range. The \nh\ (1,1) map
was obtained from the data of Torrelles \et\ (\cite{Chema94}). The triangle
shows the position of the  0.85~mm dust emission peak.  The dotted rectangle in
the HCN panel shows the area used to compute the molecular abundances in the
Ahead Core.
}
     \label{fahead} 
     \end{figure*}
%

\subsection{The Ahead Core\label{Rahead}}

We define the Ahead Core as the cold dense molecular and dust component
located  ahead of HH~2; it was first detected from observations of  \hco\
(Davis \et\ \cite{Davis90}) and \nh\ (Torrelles \et\ \cite{Chema92},
\cite{Chema94}).  Dent \et\ (\cite{Dent03}) have recently studied the submm
dust emission from the Ahead Core, which arises from a region with a size of
0.13$\times$0.09 pc$^2$. From the dust properties,  Dent \et\ (\cite{Dent03})
derive a total mass for the core of 3.8~\mo, an averaged dust temperature of
22~K and a core luminosity of  13~\lo. 

For some species (\eg\ \hco: see Fig.~\ref{fmom}), the maps of their integrated
emission over all the velocity range where the emission is detected (excluding
the high velocity component) show that the Ahead Core is not an isolated 
molecular structure but it is connected with the VLA~1 dense core.  Indeed and 
despite the BIMA primary beam attenuation, the overall morphology of the \hco\
\J{1}{0}\ agrees well with that from the \hco\ 3--2 integrated emission (Choi \&
Zhou  \cite{Choi97}).  However, we  consider the Ahead Core as the molecular
emission that is spatially coincident with  the submm dust structure detected
ahead of HH~2 (see Figure~2 of Dent \et\ \cite{Dent03}). The emission
associated  with the Ahead Core emits in the 5.2--7.8~\kms\ \vlsr\ range.
Figure~\ref{fahead} shows the averaged emission within this velocity range for
some of the molecular transitions presented in this paper as well as \nh\ (1,1)
from Torrelles \et\ (\cite{Chema94}), and the 0.85~mm dust map from Dent \et\
(\cite{Dent03}). 

The emission of the Ahead Core is traced by all the detected lines in this
work, except for the \cthd\ \JK{2}{1,2}{1}{0,1} line (which is only detected
towards the VLA 1 dense core), as can be seen in the 6, 7 and 8~\kms\ channel
maps of Figures~\ref{fcanalA}, \ref{fcanalB} and \ref{fcanalC}.  Despite the
apparent complexity of the molecular emission for the different species, we can
group them according to their morphologies within the Ahead Core.

First, \hco\ and isotopes, as well as HCN and \form\ are the only species that
trace the whole dusty structure from the Ahead Core. In particular,
Fig.~\ref{fahead} shows that there is a good spatial correspondence of these
species with the submm dust emission. \nh\ can also be included in this group,
since its emission also follows most of the dust emission (Fig.~\ref{fahead}),
especially if compared with the molecular emission from species of other groups
(see below).  In spite of the overall agreement between the emission of these
molecules and the dust, the intensity peak of the molecular emission is located
around \hhd\ knot L (only \hco\ peak coincides with that of the dust). \hco\ is
the only molecule that has emission engulfing most of \hhd, including the
brightest and highest excitation knots, H and A, where no dust is detected. 

A second group of molecules comprise SO, SO$_2$ and CH$_3$OH: they appear to be
significantly more compact than the previous species (and the dust) and their
emission is located  ahead of the brightest \hhd\ knots and with the
strongest emission spatially coincident with \hhd\ knot L (Fig.~\ref{fahead}). 
Interestingly, the emission of \metha, SO and SO$_2$ appears displaced to the
west with respect to dust peak, \ie\ apparently, they trace the face of the
dust core exposed to the \hhd\ object (see Fig.~\ref{fahead}).

%
     \begin{figure} 
     \resizebox{8.0cm}{!}{\includegraphics{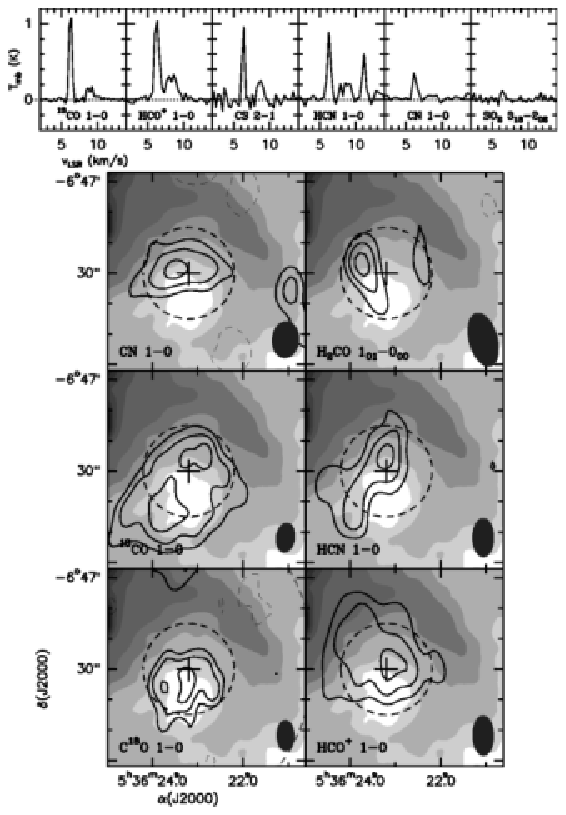}}
     \caption[]{
{\it Bottom panels:} Superposition of the grey scale [SII] image (from Curiel, 
private communication) and the \tco, \hco, \form, \cdo, CN and HCN BIMA 
contour maps  at the 6~\kms\ velocity channel of the molecular clump west of 
\hhd.  The  [SII]
image was convolved with a Gaussian with a  FWHM of 4$''$ in order  to enhance
the weak ring structure around the molecular emission.  Contours are  50, 70
and 90\% of the peak. The cross shows the position of the submm dust position
shown in Figure~2 of Dent \et\ (\cite{Dent03}).  The synthesized beam of the  
molecular maps are shown in the bottom right corner. The dashed circle shows
the area used to estimate the column densities (see \S~\ref{sawest}).
{\it Top panels:} Primary beam corrected spectra of the molecular lines 
(shown in the bottom right of the panels) taken at $\alpha (J2000) = 
5^{\rm h}36^{\rm m}23\fs2$;  $\delta (J2000) = -6\arcdeg47'30''$ and at an 
angular resolution of 30$''$. For the CN the transition shown is the $N$=1--0
$J$=3/2--1/2 $F$=5/2--3/2. The \tco\ and \hco\ line intensities have been 
scaled down a factor 3 and 2, respectively. 
}
     \label{fWclump} 
     \end{figure}
%

A third group includes CS and the CO isotopes. The CS and  \cdo\ emission and 
the bulk of the \tco\ emission from the Ahead Core arise south of the \hhd\
knot L, with little or no emission in the east part of the Ahead Core. The
peaks of emission of these molecules are located 15$''$ to 30$''$ south of the
dust peak and are roughly coincident with secondary dust peaks.

Finally, the CN emission has an inverted `L' morphology centered and peaking at
the same position as the dust. To the south of the dust peak it follows
approximately the CS emission.

\subsection{The West Core\label{Rwest}}

The West Core is the molecular structure detected west of \hhd\ at the 6~\kms\
channel of Figs.~\ref{fcanalA}, \ref{fcanalB} and \ref{fcanalC}. This core is
also traced  by the dust at submm wavelengths (Dent \et\  \cite{Dent03}).
Interestingly, very recent mid--IR observations shows that this molecular clump
is surrounded by a ring structure of hot dust (Lefloch \et\
\cite{Lefloch05}).   This hot ring shows also PAH emission (Lefloch \et\
\cite{Lefloch05}),  H$\alpha$ (Warren--Smith \& Scarrott \cite{Warren99}) and
[SII] (Fig.~\ref{fWclump}). 

The West Core is detected in \hco, \tco, \cdo, \form,  CN, HCN, CS and
marginally also in SO$_2$ and \cthd. This core is not detected in \nh\ 
(Torrelles \et\ \cite{Chema94}). Figure~\ref{fWclump} shows clearly how well 
the [SII] surrounds the molecular emission. The [SII] hole, where the molecular
emission appears, is remarkably circular with a diameter of $\sim 30''$ or
0.06~pc.  The [SII] hole is especially well traced by the CO isotopes, whereas
the CN and \hco\ are slightly shifted to the north  with respect  to the center
of this hole. The molecular line widths are quite narrow ($\Delta v \simeq
0.4$\kms) and centered at \vlsr\ of 6.3~\kms. A weaker and broader component is
also detected at \vlsr\ $=8.8$~\kms. 

%
     \begin{figure} 
    \resizebox{8.0cm}{!}{\includegraphics{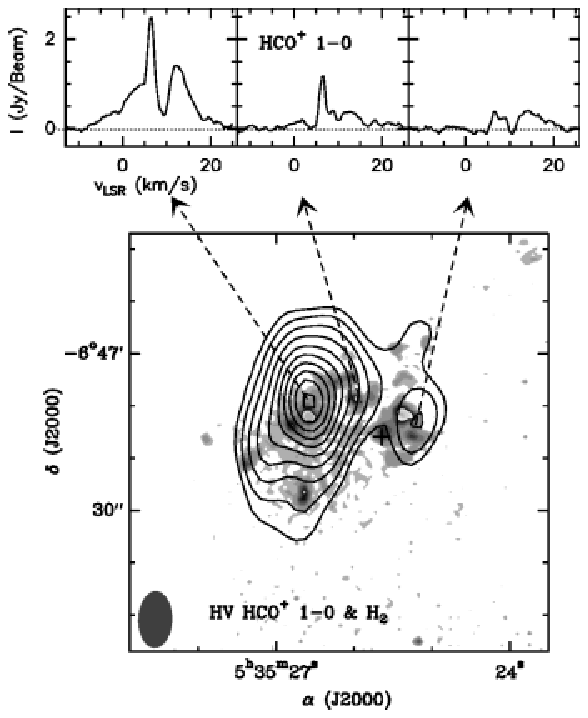}}
     \caption[]{
{\it Bottom panels}: Superposition of the grey scale near--IR\h2\ 1--0 S(1)
image (K$'$ band; Curiel, private communication) and the contour map of the
integrated high velocity \hco\ emission  over the $-$5 to 5 and 11 to 21~\kms\
\vlsr\ interval, respectively.  \hco\  contours levels are 3, 5, 8, 12, 17, 22,
27, 32, 37 and 42 times the $rms$ noise of the  map,   0.2~Jy~\kms.   The
synthesized beam is shown in the bottom left corner of each panel.  {\it Top
panel}: \hco\ \J{1}{0}\ spectra of the  high velocity  gas. The cross marks the
position of the X--ray source (Pravdo \et\ \cite{Pravdo01}).
}
     \label{fxoc} 
     \end{figure}
%
%
     \begin{figure} 
    \resizebox{8.0cm}{!}{\includegraphics{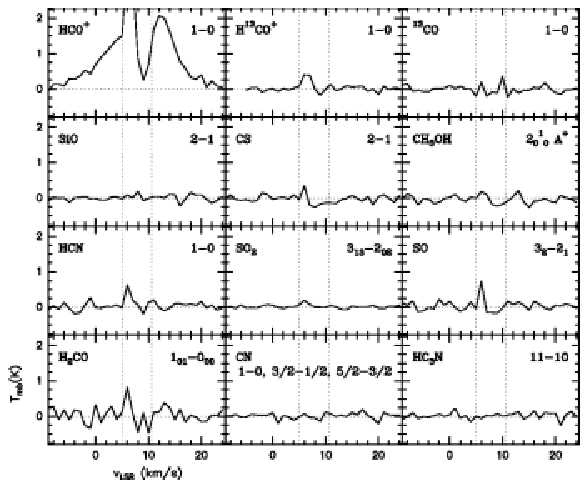}}
     \caption[]{
Spectra of the different molecular transitions at the intensity peak of the high
velocity   \hco\ \J{1}{0}\ emission ($\alpha (J2000) = 5^{\rm h}36^{\rm
m}26\fs57$;   $\delta (J2000) = -6\arcdeg47'9\farcs1$).
}
     \label{fsxoc} 
     \end{figure}
%

\subsection{High Velocity Region\label{Rxoc}}

Similarly to Dent \et\ (\cite{Dent03}), we found broad \hco\ \J{1}{0}\ emission
(\vlsr\ ranging from $-$8 to 24~\kms) spatially compact and associated with
HH~2.  Figure~\ref{fxoc} shows clearly that the high velocity \hco\ emission is
well correlated with the shock--excited near--IR 1--0 S(1)\h\ emission.  The
spatial correlation is even better with respect to the \h\ 0-0 S(2) line  (see
Fig. ~1 from Lefloch \et\ \cite{Lefloch03}):  The strongest high velocity  \hco\
emission appears located  around the \hhd\ knot E and there is also weaker 
emission near the highest excitation \hhd\ knots.  A more detailed description 
of the high velocity \hco\ around \hhd\ and its properties is given by Dent \et\
(\cite{Dent03}).

As shown in Figure~\ref{fsxoc}, and from our BIMA data, the high velocity 
emission associated with \hhd\ is detected only in the  \hco\ \J{1}{0}\ line
(the only other reported molecule with high-velocity emission is CO: \eg\
Moro--Mart{\' i}n \et\ \cite{Moro99}).  This is interesting, since species that
are usually strongly enhanced in shocked molecular clumps, such as SiO, CS,
\metha, \form\ and SO (\eg\ L1157: Bachiller \& P\'erez Guti\'errez
\cite{Bac97}) are not detected in this region.

\subsection{The overall kinematics of the dense gas\label{srcine}}

A look at the channel maps of the \hco\ emission (Fig.~\ref{fcanalA}) shows that
there is a clear velocity gradient in the dense molecular structure connecting
the Ahead Core with the VLA~1 dense core.  That is, the emission closer to the
systemic velocity (\vlsr\ $\simeq9.5$~\kms, Moro--Mart{\' i}n  \et\
\cite{Moro99}) appears located closer to the VLA~1 dense core, whereas the
emission with bluer velocities appears located closer to the Ahead Core. This is
also observed in the emission of other molecules but with a weaker
signal--to--noise ratio  (\eg\ HCN and \form).  Note that the emission from this
molecular structure from the 9 and 10~\kms\ velocity channels  of
Figures~\ref{fcanalA}, \ref{fcanalB} and \ref{fcanalC} is different for the
different molecules. Since this is the systemic velocity, these differences
could be due to high optical depths, or to other factors, such a different
excitation or chemical conditions. We will not discuss further these differences
and we will discuss only the kinematics from the \hco\ and \htco.  

\begin{figure} 
    \resizebox{\hsize}{!}{\includegraphics{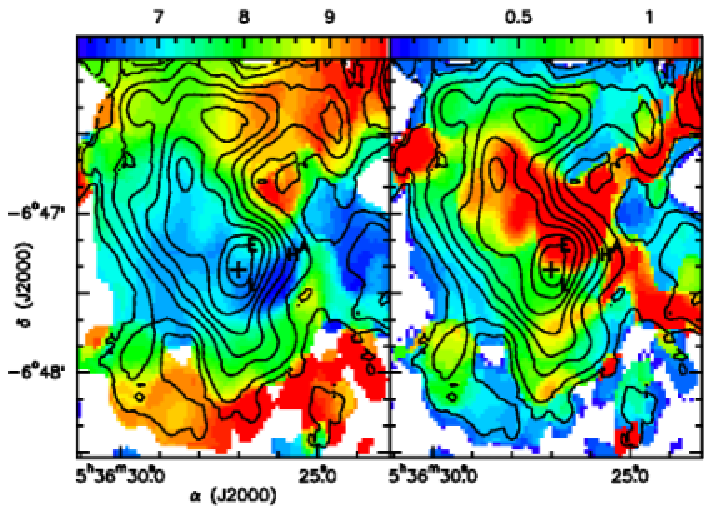}}
     \caption[]{
Superposition of the the BIMA contour maps of the zero--order moment
(integrated  emission) with the false--color image of first--order (bottom
panel) and second--order (top  panel) maps of the \hco\ \J{1}{0}. The moment
maps were obtained in the 5.35--11.26~\kms\ \vlsr\ range.  The cross shows the
position of the  0.85~mm dust emission peak.   The labels indicate the HH~2 knot
names.
}
     \label{fcinematica} 
     \end{figure}

The velocity gradient is also clearly shown in the first--order moment map of
the \hco\ (Fig.~\ref{fcinematica}).  The most blueshifted velocities,
\vlsr~$\sim 6.4$~\kms, appear close to \hhd\ knot L in the Ahead Core, and in
the West Core. South of the Ahead Core the velocity of the gas goes back to the
systemic velocity.

%
%

The second--order moment map of the \hco\ (Fig.~\ref{fcinematica}) shows that
most of the emission has a line width (FWHM) in the 0.7--1.6~\kms\ range. The
largest line widths are observed around \hhd\ knot E, where the high velocity
\hco\ is observed. Large line widths are also observed along three arms
emerging around knot E: one connecting with VLA~1 and coinciding with the
outflow axis and the other two perpendicular to this one. This region of line
width enhancement form a rotated `T--like' shape.  

\begin{figure} 
    \resizebox{\hsize}{!}{\includegraphics{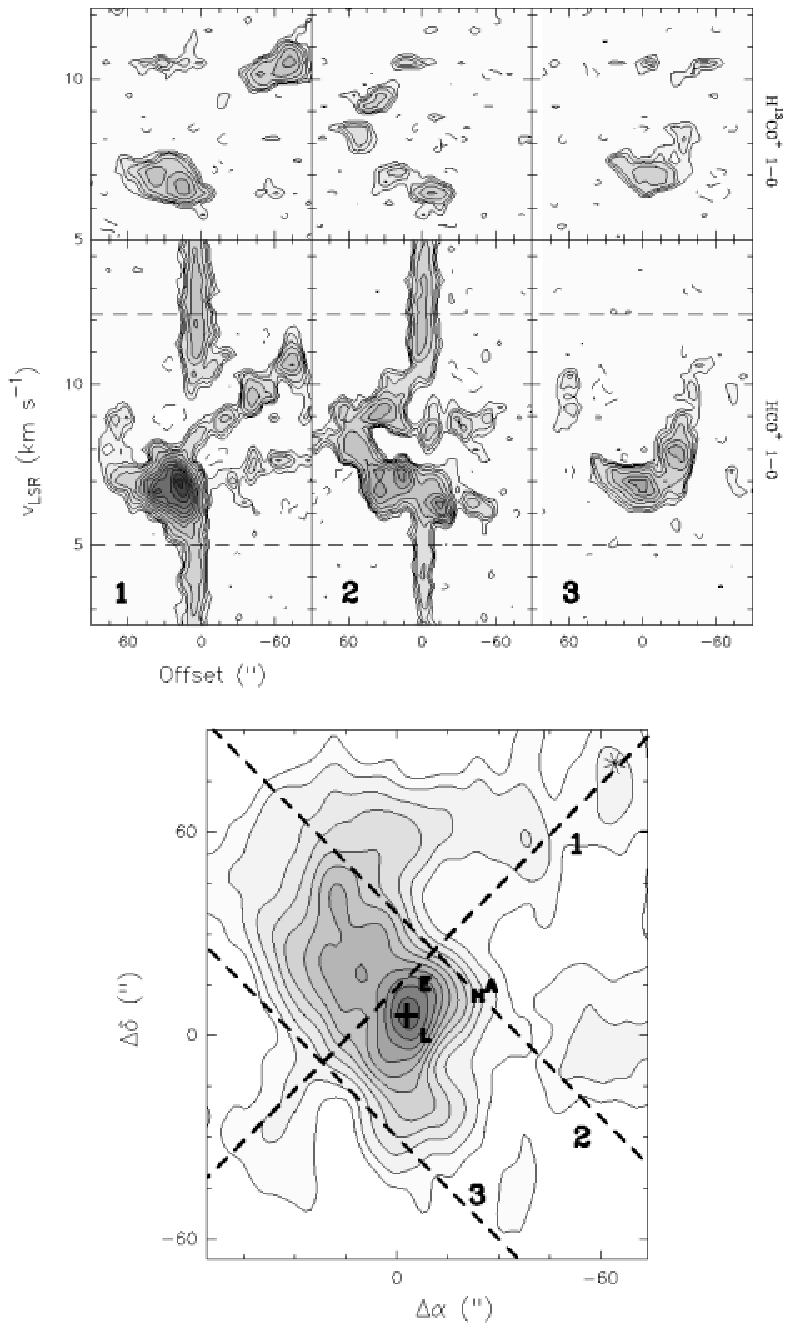}}
     \caption[]{
{\em Bottom panel:} Zero--order moment map of the \hco. The dashed lines and
numbers shows the direction of the position--velocity cuts shown in the upper 
panels.  The cross shows the position of the  0.85~mm dust emission peak.   The
labels indicates the HH~2 knot names.
{\em Upper panels:} Position--velocity maps of the \hco\ and \htco\ \J{1}{0}\ 
emission. The origin of position offsets of panels 1 and 2 are the intersection
of the dashed lines 1 and 2 in the bottom panel, while for panel 3 is the
intersection of dashed lines 1 and 3. The sign of the offsets corresponds to
the sign of $\Delta \alpha$ in the bottom panel. 
}
     \label{fpv} 
     \end{figure}

In order to see better the overall kinematics and, in particular, the region of
line width enhancement, we show in Figure~\ref{fpv} the position--velocity maps
for the \hco\ and \htco\ \J{1}{0} along three different cuts. Two of them (cuts
1 and 2) are along the direction of the line widths enhancement (see previous
paragraph). The other (cut 3) is parallel to cut 2 but further away from VLA~1
(although still within the Ahead Core).

Despite of the complexity of Figures~\ref{fcinematica} and \ref{fpv}, the
kinematics of the region can be summarized as follows:
\begin{itemize}
\item The region around \hhd\ knot E shows the largest line widths because of
the contribution from the high velocity \hco\ (this corresponds to the $0''$
offset position in the 1 and 2 cuts of Fig.\ref{fpv}).

\item One of the regions of line width enhancement (the one connecting VLA~1
with \hhd) is spatially coincident with the molecular outflow associated with
\hhd.  The \hco\ emission presents a double peak along the molecular outflow
axis (negative offset positions in cut 1 of Fig.~\ref{fpv}). The emission from
the more blueshifted peak arise from a collimated structure connecting VLA~1
with the Ahead Core (see Fig.~\ref{fesquema} and the 7~\kms\ channel of the
\hco\ of Fig.~\ref{fcanalA}). Interestingly, the CO emission from the molecular
outflow in this region is mainly redshifted with respect to the systemic
velocity (see Moro--Mart{\' i}n \et\ \cite{Moro99}), which is the reverse
situation of the \hco\ emission. Taking into account that the outflow axis is
almost on the plane of the sky, this suggests that the \hco\ is possibly
tracing the surface of interaction of the molecular outflow with the dense
molecular gas in the foreground face of the outflow lobe.

\item  Cut 3 shows that the most blueshifted velocities appear at the
intersection of the outflow axis with the Ahead Core (0$''$ offset position)
and the velocity becomes closer to the systemic velocity as the emission gets
further away from the outflow axis. The same behavior is seen in the velocity
gradient shown by the first--order moment map of Figure~\ref{fcinematica}.  Cut
2 shows a similar trend than cut 3 but, in addition, there are two other
velocity components that appears only in \hco, the high velocity component at
the 0$''$ offset position and a component at $\sim 8$~\kms. Since cut 2 follows
the border of the Ahead Core facing VLA~1, the latter component may be an
indication of interaction of the outflow with the dense core.  
\end{itemize}

%
%

\subsection{The \tco\ emission}

The \tco\ emission (\ref{fcanalC} and Fig.~\ref{fmom}) shows significantly 
more structure than the other molecular species, including \hco. This is
because of the CO low dipole moment, which allows the molecule to trace low
density  molecular gas (undetected with the higher density tracers). 
Fig.~\ref{fmom} shows that within the \tco\ field of view (which is larger than
for the other molecules because of the mosaiced observations) the emission
spreads out  almost everywhere except in a small region spatially coincident
with \hhd.  The  strongest emission surrounds \hhd.  It is interesting to note
that some of the structures traced by the high density tracers  are not well
traced or traced significantly differently by the \tco.

At the systemic velocities (9 and 10~\kms\ velocity channel maps of
Fig.~\ref{fcanalC}) the strongest \tco\ emission arises from a clear V--shaped
structure, with its vertex located near VLA~1, facing and surrounding \hhd. 
The northern part of this structure corresponds approximately to the high
density molecular gas that connects the Ahead Core with the VLA~1 dense core.
Given the direction of the outflow (from VLA~1 to \hhd), the V--shaped
morphology seems to trace the walls of a cavity created by the HH~1--2
outflow.  Indeed, within this cavity and connecting VLA~1 and \hhd\ there is a
highly collimated molecular outflow (Moro--Mart{\' i}n  \et\ \cite{Moro99}).

\section{Analysis}

In order to compare properly the different species and derive the temperature
and column densities, we first made maps of the different transitions at the
same angular resolution.  This angular resolution was set to $15''\times9''$, 
$PA\simeq 0\arcdeg$.

\subsection{The Ahead Core}

The physical parameters for the Ahead Core were derived from the maps of the
integrated emission within the  5.2--7.8~\kms\ velocity  range.  

\subsubsection{Optical depth of the \hco\ and HCN 1--0 lines\label{stau}}

The \hco\ 1--0 optical depth was derived from the ratio of the \htco\ and  \hco\
integrated emission maps, corrected for the primary beam response. The highest 
values of the \hco\ 1--0 optical depth within the FWHM of the primary beam  are
found  south of \hhd, where the optical depth reaches up to $\sim 20$.  The
optical depth smoothly decreases to 2--4 east and northeast of \hhd.

The \htcn\ 1--0 is not detected, so only upper limits can be derived for the
HCN 1--0 optical depth.  Since the HCN is quite extended, channel maps of both
isotopes were smoothed to an angular resolution of 20$''$.  From these maps we
estimate that $\tau_{\rm HCN } \la 10$. This implies that the observed HCN
column density derived in Paper I could be underestimated by a factor of $\la
10$.  In fact, from Paper II, the best matching model (B15) at $\sim$ 100 yrs
gave HCN abundances ranging from $\sim$ 2$\times$10$^{11}$ cm$^{-2}$ at 1 mag
to $\sim$ 6$\times$10$^{16}$ cm$^{-2}$ at 5 mags; for a 3 mags gas we have a
theoretical column density of 2.7$\times$10$^{13}$ which is indeed only a
factor of 10 higher than the observed column density of 4$\times$10$^{12}$
cm$^{-2}$.

\subsubsection{Mean physical parameters\label{samean}}

%
     \begin{table}
     \caption[]{Mean column densities and abundances in the Ahead Core}
     \label{tahead}
     \[
\begin{tabular}{lll}
     \noalign{\smallskip}
     \hline
     \noalign{\smallskip}   
\multicolumn{1}{c}{} &
\multicolumn{1}{c}{$N$[mol]} 
\\
\multicolumn{1}{c}{Molecule} &
\multicolumn{1}{c}{(\cmd)} &
\multicolumn{1}{c}{$X$[mol]$^b$} 
\\
     \noalign{\smallskip}
     \hline
     \noalign{\smallskip}
CO$^a$      & 2.9\N{17}& 3.2\N{-5}  \\
H$_2$CO & 3.0\N{13}& 3.4\N{-9}  \\
\hco\   & 2.9\N{13}& 3.2\N{-9}  \\
\metha\ & 1.4\N{13}& 1.5\N{-9}  \\
SO      & 6.7\N{12}& 7.4\N{-10} \\
HCN     & 1.5\N{12}& 1.6\N{-10} \\
SO$_2$  & 1.2\N{12}& 1.3\N{-10} \\
CS      & 9.2\N{11}& 1.0\N{-10} \\
CN      & 6.7\N{11}& 7.4\N{-11} \\
DCO$^+$ & 2.2\N{11}& 2.4\N{-11}  \\
     \noalign{\smallskip}
     \hline
     \end{tabular}
     \]
     \begin{list}{}{}
\item[$^a$] 
Here after the CO column densities are derived adopting a 
$^{12}$CO to \tco\ ratio of 63 (Langer \& Penzias  \cite{Langer93}). 
\item[$^b$] $X$[mol] is the abundance with respect to H$_{2}$
     \end{list}
    \end{table}
%

The total mass of the Ahead Core is 3.8$\pm$0.4~\mo, as derived from submm dust
observations (Dent \et\ \cite{Dent03}). To estimate the mean column densities 
and molecular abundances of the observed species over the whole Ahead Core
(using the mass derived from the dust), we integrated the emission of the
different species within the  5.2--7.8~\kms\ velocity range and over a box of
$1\farcm0\times1\farcm5$ that encompasses the dust emission ahead of \hhd\
shown by Dent \et\ (\cite{Dent03}).  The column densities were estimated
assuming  an excitation temperature of 10~K. Table~\ref{tahead} shows the
estimated absolute molecular abundances averaged over the whole Ahead Core. 
The relative abundances with respect to the CO are a factor $\sim$2--10 lower
than the derived from Paper I. Note however that the relative abundances given
in Paper I were obtained for a FWHM beam of 30$''$  at the SO peak,  whereas
the values derived in Table~\ref{tahead} were obtained over a larger region. 
This difference can be interpreted as arising from chemical gradients within
the Ahead Core.

\subsubsection{Gradients within the Ahead Core\label{sagrad}}

%
     \begin{figure} 
    \resizebox{6.5cm}{!}{\includegraphics{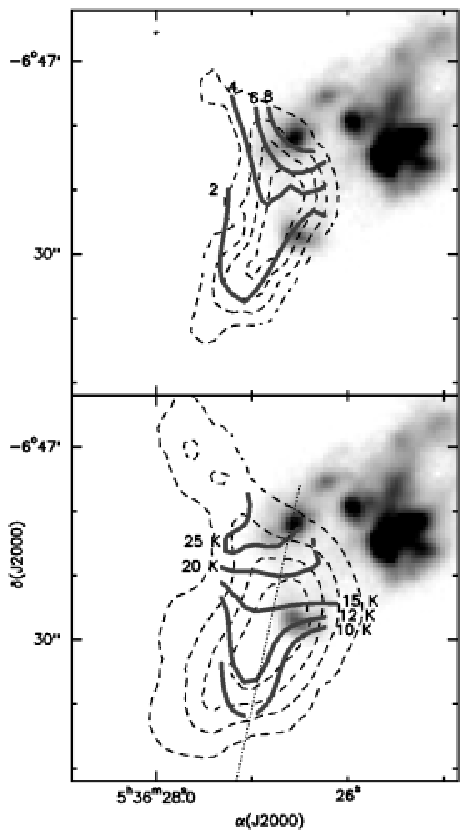}}
     \caption[]{
{\it Top panel}:  Superposition of the $10^{-5}\, X[{\rm SO_2}]/X[{\rm CO}]$ 
map  (thick grey contours),  the zero--order moment map of the SO$_2$
\JK{3}{1,3}{2}{0,2}\ emission (dashed contours) and [SII] emission (grey scale
image). 
{\it Bottom panel}: Superposition of the temperature map (thick grey contours)
derived  from the SO \JK{2}{3}{1}{2}\ to SO \JK{3}{2}{2}{1}\ ratio,  the
zero--order  moment map of the SO \JK{3}{2}{2}{1}\ emission (dashed contours)
and the [SII] emission (grey scale image). The dotted line shows the slice done
for Figure~\ref{ftall}.
}
     \label{ftemp} 
     \end{figure}
%

We computed the excitation temperature of the SO for the  \JK{3}{2}{2}{1}\ and
\JK{2}{3}{1}{2}\ lines.  By using these two  transitions the excitation
temperature can be expressed as: 
\begin{equation}
T_{\rm ex} \simeq -11.7 \, \left ({\rm log} ( 1.77 \,
\frac{\int I ({\rm SO \, 2_3-1_2}) \, dv}{\int I({\rm SO \, 3_2-2_1})\, dv} 
)  \right )^{-1}
\end{equation}
where $I$ is the line intensity (in K). The resulting  excitation temperature
map is  shown in Figure~\ref{ftemp}; there is a gradient, with 
increasing  excitation  temperatures towards \hhd.  

Column  density maps for  SO$_2$ and \tco\ were derived assuming an  excitation
temperature of 10~K.  Figure~\ref{ftemp} shows  the map of the SO$_2$ relative
abundance with respect to the CO.  The map shows a gradient in the  SO$_2$
relative abundance,  increasing towards \hhd, having the highest value close to
the \hhd\ knot E.   Note that, if instead of using a constant $T_{\rm ex}$, we
use the values  derived from the SO line ratio analysis (equation 1) the
results do not significantly change. 

%
     \begin{figure} 
     \resizebox{\hsize}{!}{\includegraphics{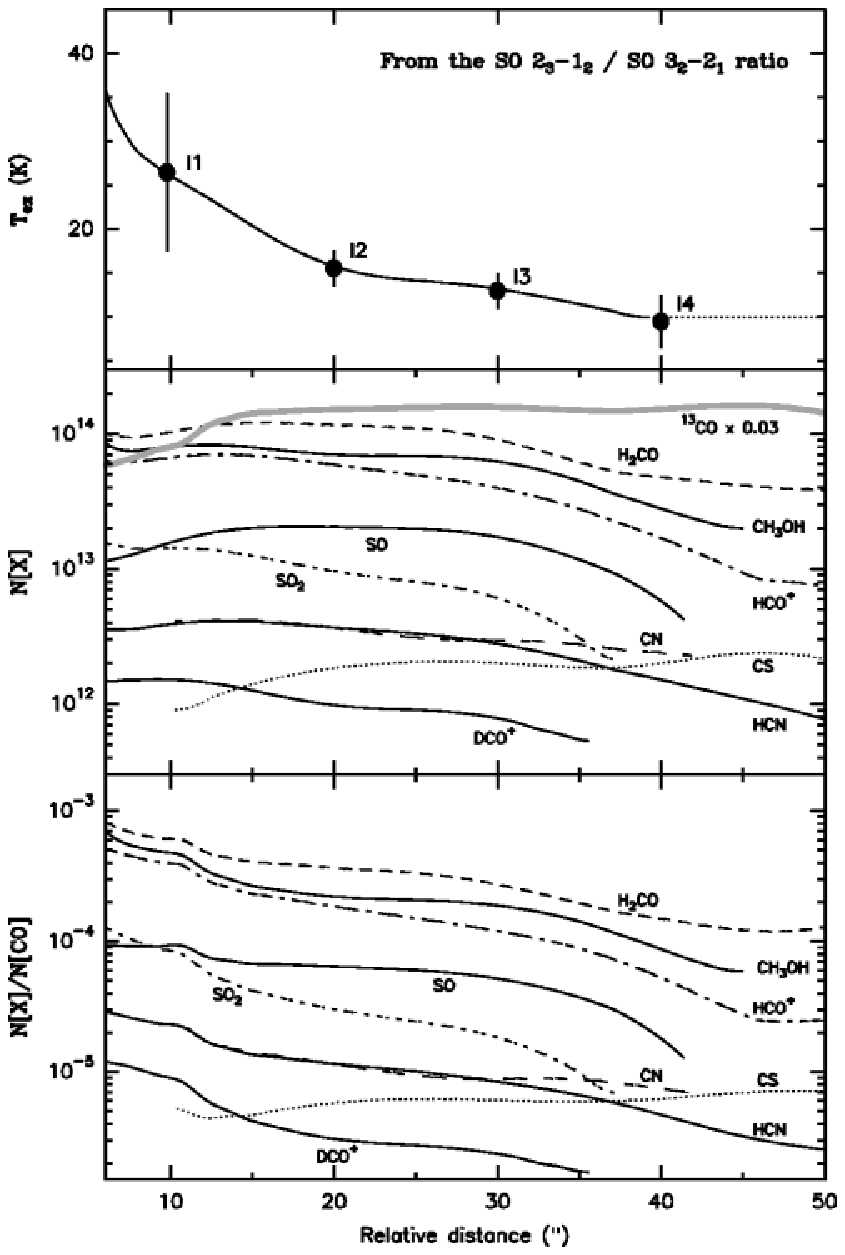}}
     \caption[]{
Slices of the SO excitation temperature (top panel), column density (central
panel) and  relative abundance  with respect to the CO (bottom panel) as a
function of the  distance to \hhd.  The first position in the plot coincides
with the position of the \hhd\  knot E. The slice has 0 position  
$\alpha (J2000) = 5^{\rm h}36^{\rm m}26\fs5$;   
$\delta (J2000) = -6\arcdeg47'06'\farcs0$ and the position angle of the slice
is PA=169\arcdeg. Note that the difference in column density with Figure~1 from
Paper II are  because in paper II we used a constant $\T{ex}$ whereas in this
paper we used the value derived from the SO line ratio. In the case of \hco,
the new observations of \htco\ allowed to correct for optical depth.
}
     \label{ftall} 
     \end{figure}
%

%
\begin{table}
\caption{Column densities for selected positions along the slice in the 
SO$_2$ Clump}
\label{tilumina}
\begin{tabular}{l@{\hspace{0cm}}c@{\hspace{2mm}}c@{\hspace{2mm}}c@{\hspace{2mm}}c@{\hspace{2mm}}r}
     \noalign{\smallskip}
     \hline
     \noalign{\smallskip}   
\multicolumn{1}{l}{Position} &
\multicolumn{1}{c}{I1} &
\multicolumn{1}{c}{I2} &
\multicolumn{1}{c}{I3} &
\multicolumn{1}{c}{I4} &
\multicolumn{1}{c}{} 
\\
     \noalign{\smallskip}
     \hline
     \noalign{\smallskip}
T$_{ex}$   (K)$^a$  & 27$\pm$9 & 16$\pm$2 & 13$\pm$2 & 9$\pm$3 & \\
$d_{\rm slice}$ ($''$)$^b$& 10 & 20 & 30 & 40 & \\
$d_{\rm HH 2}$ ($''$)$^c$& 16 & 22 & 29 & 39 & \\
     \noalign{\smallskip}
     \hline
     \noalign{\smallskip}   
\multicolumn{1}{l}{} &
\multicolumn{4}{c}{$N$[mol]} &
\multicolumn{1}{r}{$\!\!\!\!\!\!\!\!\!X_{\rm I1}$} 
\\
\multicolumn{1}{l}{Molecule} &
\multicolumn{4}{c}{(\cmd)} &
\multicolumn{1}{r}{$\!\!\!\!\!\!\!\!\!\!\!\!/ X_{\rm I4}$} 
\\
     \noalign{\smallskip}
     \hline
     \noalign{\smallskip}
CO      & 1.7\N{17} & 3.2\N{17} & 3.3\N{17} & 3.2\N{17} &   1   \\
H$_2$CO & 1.0\N{14} & 1.2\N{14} & 9.0\N{13} & 4.8\N{13} & 4$\pm$1 \\
\hco\   & 6.7\N{13} & 6.0\N{13} & 4.0\N{13} & 1.8\N{13} & 8$\pm$1 \\
\metha\ & 8.0\N{13} & 7.1\N{13} & 6.2\N{13} & 2.8\N{13} & 6$\pm$2 \\
SO      & 1.6\N{13} & 2.1\N{13} & 1.7\N{13} & 5.7\N{12} & 5$\pm$1 \\
HCN     & 3.9\N{12} & 3.7\N{12} & 2.8\N{12} & 1.5\N{12} & 5$\pm$1 \\
SO$_2$  & 1.4\N{13} & 9.7\N{12} & 6.0\N{12} &$<$2\N{12} & $\ga$12 \\
CS      & 9.1\N{11} & 1.8\N{12} & 2.0\N{12} & 2.0\N{12} &0.8$\pm$0.3\\
CN      & 4.1\N{12} & 3.7\N{12} & 3.0\N{12} & 2.4\N{12} & 3$\pm$2 \\
DCO$^+$ & 1.5\N{12} & 9.9\N{11} & 7.8\N{11} &$<$5\N{11} &$\ga5$ \\
\hline
\end{tabular}
     \begin{list}{}{}
\item[$^a$] Error bars are at 1-$\sigma$
\item[$^b$] Relative distance in the slice (see Fig.~\ref{ftall})
\item[$^c$] Distance to \hhd\ knot H 
\end{list}
\end{table}
%

Given the SO excitation temperature and SO$_{2}$  relative abundance gradient
towards \hhd\ shown in Figure~\ref{ftemp}, we measured the integrated emission
along the direction of this gradient, which approximately coincides with the SO
and SO$_2$ major axis.  From these slices we computed: (1) The  excitation
temperature for the SO  \JK{3}{2}{2}{1}\ and  \JK{2}{3}{1}{2}\ lines (see first
paragraph of this section).  (2) The column density of the different molecular
species assuming an excitation temperature equal to the SO  excitation 
temperature. For \hco,  its column  density was corrected by taking into account
$\tau_{\rm HCO^+}$ wherever \htco\  was detected, otherwise $\tau _{\rm HCO^+} =
6.0$  was adopted (which is the lowest  $\tau_{\rm HCO^+}$ value measured in the
slice). (3) The relative abundances  with respect to CO. 

Figure~\ref{ftall} shows the three parameters, SO excitation temperature, 
column density and abundance, along the slice. Table~\ref{tilumina} gives the
SO excitation temperature, column density at  four positions along the slice, as
well as the abundance increase factor at the first and last positions. It is
clear from this figure that  the relative abundance enhancement towards  \hhd\
is observed in all molecular species except CS, which shows a relative abundance
almost constant along the slice.  \hco\ and SO$_2$  are the species that have
the highest enhancement.

\begin{figure} 
     \resizebox{\hsize}{!}{\includegraphics{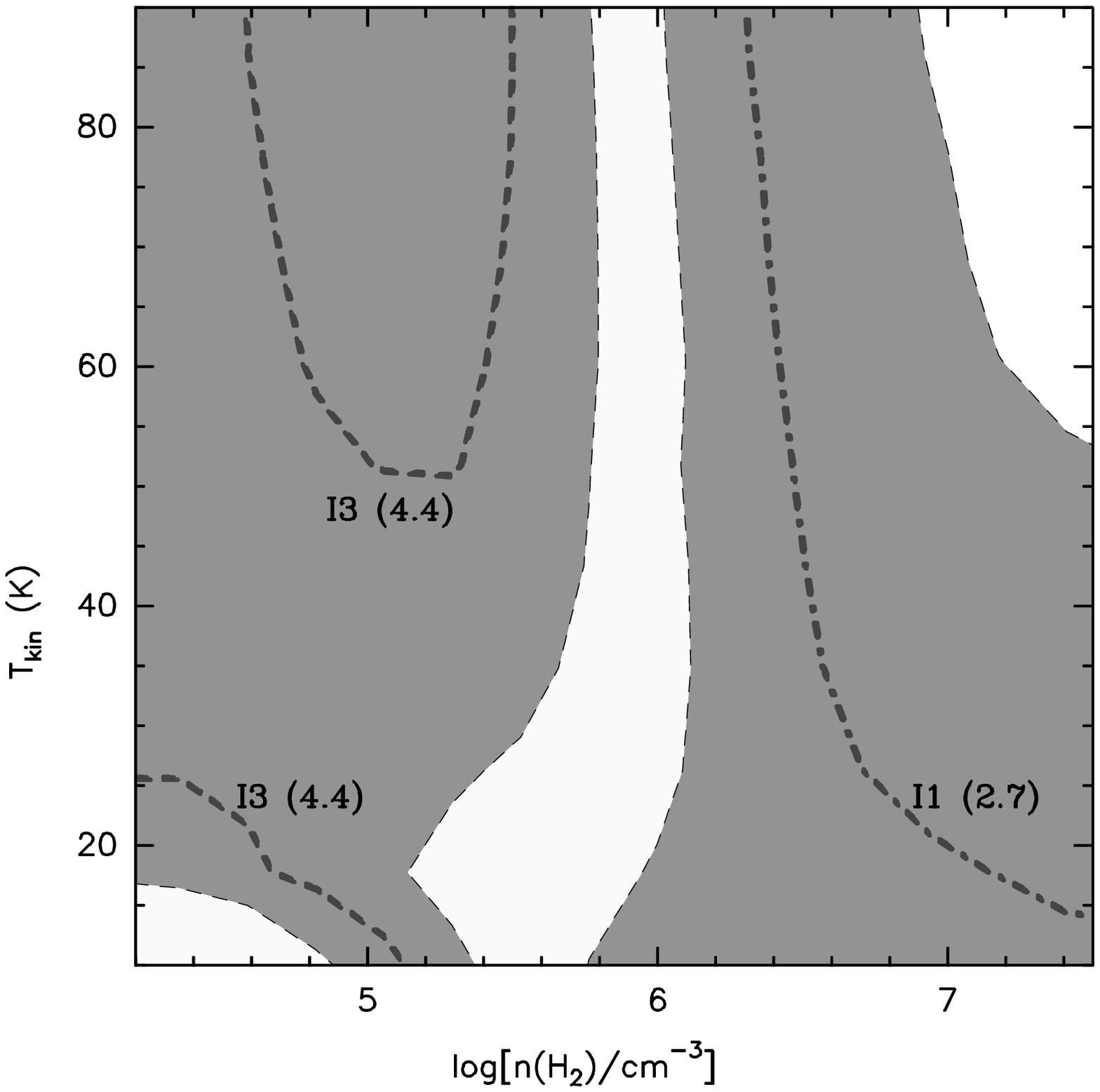}}
     \caption[]{
Set of RADEX solutions in the $T_{\rm kin}$--$n$(H$_{2}$) plane, for the
observed  SO \JK{3}{2}{2}{1}\ / \JK{2}{3}{1}{2}\ line intensity ratios (assuming
a SO column density of 2\N{14}~\cmd). The thick dashed lines is the contour  for
the ratio at  position I3 (4.4) and the thick dotted-dashed  line is the contour
for the ratio at position I1 (2.7).  The grey area shows the area within a $\pm
1\sigma$.
}
     \label{fradex} 
     \end{figure}

In order to interpret the excitation temperature enhancement  we used the RADEX
package to fit the observed SO \JK{3}{2}{2}{1}\ / \JK{2}{3}{1}{2}\ line ratios 
at the positions I1 (2.7$\pm0.4$) and I3 (4.4$\pm$0.6) given in
Table~\ref{tilumina}.  RADEX  is a non--LTE molecular radiative transfer in an
isothermal homogeneous medium  (Sch{\" o}ier \et\ \cite{Schoier05}).   In
Figure~\ref{fradex} we show a set of the RADEX solutions in the $T_{\rm
kin}$--$n$(H$_{2}$) plane for the observed line ratios. The set of solutions in
Fig.~\ref{fradex} that satisfy these averaged values indicates that the density
in  I1 ($\ga 10^6$~\cmt) is higher than in I3 ($\sim 10^5$~\cmt),  although the
temperature cannot be constrained. Further observations of higher excitation
lines at high angular resolution will help to better constrain the density
enhancement and to determine the temperature. 

%
%

\subsection{West Core\label{sawest}}

We first corrected the channel maps from the BIMA primary beam response  and
then convolved them with a Gaussian with a FWHM of 30$''$ in order to optimize
the signal--to--noise ratio.  The column densities were estimated from the
resulting map at the position of the dust peak (see Figure~\ref{fWclump}) and
for the \vlsr\ of 6.3~\kms\ component. Since we do not have information about
the temperature of the molecular gas (no SO line is detected towards this core),
we assumed an excitation temperature of 10~K. The narrow line widths of this
velocity component indicate that the molecular gas is not shocked and,
therefore, the temperature is probably low: even in the case that the West Core 
is in a similar situation as the  SO$_2$  Clump (see \S~\ref{sdwest}), the
region with high temperatures would be probably compact.  Fig.~2 of Dent \et\
(\cite{Dent03}) shows that the 850~$\mu$m dust emission peak towards the West
Core is $\sim90$~\mjy. Assuming that the dust temperature and opacity is similar
to the Ahead Core, then the mean gas column density in the West Core at the dust
peak is $N($H$_{2})\simeq 1$\N{22}~\cmd.  The beam averaged \tco\ column density
at the dust peak and at an $14''$~angular resolution (the same as the dust map)
is 2.8\N{15}~\cmd, which gives an abundance of $X$[CO]$\simeq1.8$\N{-5}.
Table~\ref{twest} shows the derived column densities and relative abundances for
the 30$''$ smoothed spectra (adopting the aforementioned $X$[CO] value). From
the \tco\ emission, we derive a mass of $\sim 0.5$~\mo\ for the West Core.

%
     \begin{table}
     \caption[]{Column densities and abundances of the West Core}
     \label{twest}
     \[
     \begin{tabular}{lrrr}
     \noalign{\smallskip}
     \hline
     \noalign{\smallskip}   
\multicolumn{1}{c}{} &
\multicolumn{1}{c}{$N$[mol]} &
\multicolumn{1}{c}{$X[$mol$]/$} &
\multicolumn{1}{c}{} 
\\
\multicolumn{1}{l}{Molecule} &
\multicolumn{1}{c}{(\cmd)} &
\multicolumn{1}{c}{$X[$CO$]$} &
\multicolumn{1}{c}{$X[$mol]$^a$} 
\\
     \noalign{\smallskip}
     \hline
CO    & 1.7\N{17}   &  1.0      & 1.8\N{-5}    \\
H$_2$CO   & 1.3\N{13}   &  8\N{-5}  & 1.4\N{-9}    \\
\hco\     & 6.8\N{12}   &  4\N{-5}  & 7.2\N{-10}   \\
\metha    & \mq2\N{13}  & \mq1\N{-4}&\mq2\N{-9}    \\
SO        & \mq2\N{12}  & \mq1\N{-5}&\mq2\N{-10}   \\
HCN       & 1.6\N{12}   &  9\N{-6}  & 1.7\N{-10}   \\
SO$_2$    &$\sim$2\N{12}&  1\N{-5}  &$\sim$2\N{-10}\\
CS        & 1.6\N{12}   &  9\N{-6}  & 1.7\N{-10}   \\
CN        & 3.3\N{12}   &  2\N{-5}  & 3.5\N{-10}   \\
DCO$^+$   &$\sim$2\N{11}&  1\N{-6}  &$\sim$2\N{-11}\\ 
\cthd     &$\sim$7\N{11}&  4\N{-6}  &$\sim$7\N{-11}\\
     \noalign{\smallskip}
     \hline
     \end{tabular}
     \]
     \begin{list}{}{}
\item[$^a$] $X$[mol] is the abundance with respect to H$_{2}$
     \end{list}
    \end{table}
%

\subsection{The High Velocity Region}

The column density for the \hco\ and the upper limits (at $3\sigma$) for the 
rest of the molecules were estimated at the intensity peak of the high velocity
\hco\ emission (see Fig.~\ref{fsxoc}), which is $\sim 3''$ northwest of the
\hhd\ knot E. The column densities are estimated from the integrated emission
over the $-5$ to 5 and 11 to 21~\kms\ velocity ranges, avoiding the contribution
from the ambient gas of the Ahead Core (and the  SO$_2$ Clump), and assuming an
excitation temperature of 30~K. Table~\ref{txoc} shows the column densities and
the relative abundances of several molecules  with respect to \hco, and for
comparison we also give typical relative abundances in shocked regions of
molecular outflows  driven by low mass protostars.  It is clear from this table
that the \hco\ abundance is strongly enhanced or, alternatively, that the other
molecules are severely  depleted in the  High Velocity Region (see
\S~\ref{sdxoc}).

%
     \begin{table}
     \caption[]{Column densities and abundances in the High Velocity Region}
     \label{txoc}
     \[
\begin{tabular}{l@{\hspace{0cm}}r@{\hspace{2mm}}r@{\hspace{2mm}}r@{\hspace{1mm}}r@{\hspace{2mm}}r}
     \noalign{\smallskip}
     \hline
     \noalign{\smallskip}   
\multicolumn{2}{c}{} &
\multicolumn{4}{c}{N(mol.)/N(HCO$^+$)} 
\\
\cline{3-6}
\multicolumn{1}{c}{} &
\multicolumn{1}{c}{$\!\!N$(mol)} &
\multicolumn{1}{c}{} &
\multicolumn{1}{c}{Shocked} &
\multicolumn{1}{c}{Model} &
\multicolumn{1}{c}{Model} 
\\
\multicolumn{1}{c}{Molecule} &
\multicolumn{1}{c}{(\cmd)} &
\multicolumn{1}{c}{HH 2} &
\multicolumn{1}{c}{Regions$^{a}$} &
\multicolumn{1}{c}{A$^b$} &
\multicolumn{1}{c}{B$^b$} 
\\
     \noalign{\smallskip}
     \hline
     \noalign{\smallskip}
\hco   & 4.1\N{13} & 1      &   1        & 1      & 1      \\
CO     &$<$1\N{17} &$<$3000 &2000--3\N{5}& 2\N{4} & 1\N{4} \\
\form  &$<$3\N{14} &$<$8    &   4--19    & 6.5      & 3.9  \\
\metha &$<$3\N{14} &$<$7    &  70--2000  & 2\N{-4}& 4\N{-4}\\
SO     &$<$3\N{13} &$<$0.7  &   4--37    & 0.05   & 0.14   \\
HCN    &$<$4\N{12} &$<$0.09 &   2--18    & 200    & 500    \\
SO$_2$ &$<$3\N{13} &$<$0.8  &   4--19    & 3\N{-4}& 0.02   \\
CS     &$<$7\N{12} &$<$0.2  &   4--13    & 0.17   & 0.70   \\
CN     &$<$3\N{13} &$<$0.7  & 1.4--5     & 2.6    & 250    \\
HC$_3$N&$<$4\N{12} &$<$0.1  & 0.2--0.8   & 4\N{-4}& 9\N{-4}\\
SiO    &$<$4\N{12} &$<$0.09 & 0.6--40    &\nodata &\nodata \\
     \noalign{\smallskip}
     \hline
     \end{tabular}
     \]
     \begin{list}{}{}
\item[$^a$] Range of values got from the molecular outflows in L1157 (Bachiller
\& P\'erez Guti\'errez \cite{Bac97}),  NGC 1333 IRAS2A (J{\o}rgensen \et\
\cite{Jorgensen04a}) and  BHR71 (Garay et al.\cite{Garay98})
\item[$^b$] Theoretical ratios derived for shocked molecular gas under a strong
UV fields: see more details in \S~\ref{sdxoc} 
    \end{list}
    \end{table}
%

\section{Discussion}

\begin{figure} 
    \resizebox{\hsize}{!}{\includegraphics{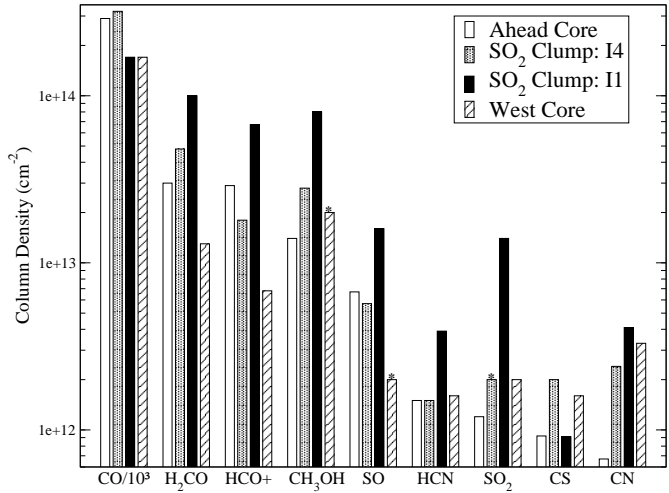}}
     \caption[]{
Comparison of the column densities of several molecular species for the Ahead
Core, the West Core and the positions I1 and I4 of the SO$_{2}$ clump. Position
I1 is closer to \hhd\ than I4 (see Fig.~\ref{ftall} and Table~\ref{tilumina}).
Asterisks on the bar means upper limit of the column density.
}
     \label{fregions} 
     \end{figure}

In this section, we attempt to provide possible explanations for the origin and
structure of the observed molecular  emissions in the frame of UV illumination
models.  We do not in any way attempt to model any of the regions in  great
detail.  From now on we will define the  SO$_2$ Clump as the molecular
component within the Ahead Core traced by SO, SO$_2$ and \metha\ and whose
chemical  properties were studied in Papers I and II (see Fig.~2 from Paper I).
The  SO$_2$  Clump  appears located in the region of the Ahead Core facing
\hhd\ (and thus is the molecular region in the Ahead Core more exposed to the
UV radiation). Therefore, we split the discussion of the  SO$_2$  Clump and
Ahead Core in two different subsections.   Figure~\ref{fregions} shows the
column densities for the Ahead Core,  SO$_{2}$ Clump and the West Core.

%
%

\subsection{The SO$_2$ Clump}

In \S~\ref{sagrad} we found an SO excitation temperature and relative molecular
abundance enhancement within the SO$_2$ Clump towards \hhd\ (see
Figures~\ref{ftemp} and \ref{ftall}). The excitation temperature enhancement
indicates  a density increase towards \hhd, which may suggest compression from
the wind.   Although the position at which the  molecular enhancement occurs
coincides  very well with \hhd\ knot E, the most likely source of UV radiation
are the high  excitation knots H and A (\eg\ B\"ohm \et\ \cite{Boehm92}).   This
could be due to a projection effect: \ie\ the molecular gas spatially coinciding
with knot E could be the closest to knot H and A.

It is possible that the photoelectric heating mechanism due to the strong UV
radiation could cause an increase in temperature in the molecular gas.   We
investigated this scenario  by running a PDR model that self--consistently 
computes the temperature at the edge of the clump (see Papadopoulos, Thi \&
Viti \cite{Papa02} and  Bell \et\ \cite{Bell05}). We found that with a 
radiation field of $\sim$40 Habing (Molinari \& Noriega--Crespo 
\cite{Molinari02}; 1 Habing = 1.6$\times$10$^{-3}$ erg cm$^{-2}$~s$^{-1}$) 
a temperature of  50~K  is reached  1.0~mags into the clump  for densities
between 10$^5$ and 10$^6$~cm$^{-3}$.

As already mentioned  there is a definite gradient in both  density and
abundance.  In light of this result, and recalling the conclusions of Paper II
it is clear that the  molecular enhancements found in the  SO$_2$ Clump do
$not$ come from a  single density component gas. We have therefore computed the
observed column densities and the temperatures at four positions along the
slice of Figure~\ref{ftall}, see Table~\ref{tilumina}. In this table we also
give the distance between the positions selected and the \hhd\ knot H (which
together with knot A are the likely source of the UV radiation).  We then ran
multi--density components models, using the chemical model employed in Paper II
and revised visual  extinctions.  The physical and chemical parameters of these
models were as those in B2 and in B7 in Paper II, but we used a number density
structure with a minimum and  maximum number  density of respectively
1$\times$10$^5$ and 3$\times$10$^5$ cm$^{-3}$. The maximum  number density is
as in B2 and B7 of Paper  II, while the minimum density is determined by
applying the density law derived by Tafalla et al. (\cite{Tafalla02}) for
starless cores. Since in our models  the visual extinction is a function of the
column density and therefore of the density, the visual extinction at any point
in the modelled clump has changed from B2 and B7 accordingly.

We find that it is possible to match, within a factor of 5, all the species
everywhere if the visual extinction (along the line of sight to the source) is
$\sim 3$~mags for all positions, and for ages $<$ 1000 yrs  ($after$ it starts
being irradiated).  Since we know the distance of the emitting region to the UV
source ($d_{HH2}$ in Table 4), a constant visual magnitude suggests one of the
following three scenarios:  (1) a decreasing (from \hhd) density structure with
densities of 1.4$\times10^5$ cm$^{-3}$, 7.4$\times10^4$ cm$^{-3}$,
5$\times10^4$ cm$^{-3}$ and 3.5$\times10^4$ cm$^{-3}$ for positions I1, I2, I3
and I4, respectively (see Table~\ref{tilumina});  (2) a clumpy structure, where
the density and size of each clump is constant (and hence  the column density
within each clump along the line of the source),  and the difference in the
observed column densities is simply due to the fact that as the radiation field
crosses several clumps, it gets attenuated. This implies that the four
positions correspond more or less to four clumps along the same line with
respect to the source;  (3) a third scenario may involve a more complicated
picture where the line  joining knots A, H and the  SO$_2$ Clump is not
perpendicular to the  line of sight, hence the radiation field may be impinging
on the  SO$_2$  Clump from the top  (or the bottom)  along the line of sight to
the observer.  Probably the most likely scenario is a combination of the first
and third  scenarios. The first one agrees with the density increase towards
\hhd\ derived  from the SO line ratios. The third one may agree with the
distribution  of the SO, SO$_{2}$ and \metha\ with respect to the dust: these
molecules  appear located in the exposing face (with respect to \hhd) of the
strongest 0.85~mm  dust peak from the Ahead Core. In this case, the  SO$_2$ 
Clump may not be a true clump, but just the illuminated face of a slightly
larger clump, whose center is located at the position of the 0.85~mm dust
peak.  

On the basis of this model,  we conclude that the  \hhd\  is the source of
`illumination'  and that the  SO$_2$ Clump must have substructure  (a density
increase towards \hhd, as a result of compression from the wind)  within it
(re--iterating a main conclusion of Paper II) and that this accounts for the
chemical variation across the core. However, it is not possible to deduce from
these studies whether the  SO$_2$ Clump is a core distinct from the Ahead Core
or is simply a face of Ahead Core strongly affected by the radiation from HH2.

\subsection{The Ahead Core\label{sdahead}}

The Ahead Core clearly shows some  molecular abundance differences with
respect  to a typical starless dense core: for example, comparing the Ahead
Core abundances with  those from L1544 and L1689B (J{\o}rgensen, Sch{\" o}ier,
\& van Dishoeck  \cite{Jorgensen04b}), both CO and HCO$^+$ are slightly
enhanced (a factor $\sim 2$--8), whereas the CS is depleted by a factor 9--26.
We now explore whether,  as  for the  SO$_2$ Clump, the abundances can be
reproduced as a consequence of the UV field from the \hhd\ shock  impinging on
the gas and dust of the core. 

Excluding the  SO$_2$  Clump, the Ahead Core has a large size, $\sim$ 0.1~pc,
hence it is reasonable to assume that the species we are detecting may all be
tracing different density components in this core. We used  the models in Paper
II to determine whether the chemistry of the core can be explained by a weak
radiation field, because further away, that is impinging on a fairly dense
core.  We find that the results at early times ($<$ 500 yrs) of models from
both Grid A and B from Paper II may match the observations fairly well  (within
a factor of 5, and with a radiation field of $\sim$ 10--20 Habing)  if one
assumes that i) different densities along the line of the source are present
and that ii) different molecules trace different components. For example, CS
and CO emission  seem to come mainly  from the southern part of the Ahead Core,
correlating well with a secondary dust peak, while HCN, \form\ and \hco\ trace
the whole core. These considerations suggest that the Ahead Core may not be an
homogeneous and roughly spherical core but it may be composed of small clumps
similar to those found in L673 (Morata, Girart \& Estalella \cite{Morata05}).
These BIMA observations reveal small--scale clumps (typically a few $\times$
10$^{-2}$ pc) in lines of HCO$^+$, N$_2$H$^+$, and CS. However, the maps of
these species did not coincide; this is probably a consequence of the
time--dependence of the chemistry (Garrod \et\ \cite{Garrod05}).  A detailed
modelling of this region, taking into considerations these assumption should be
attempted.

\subsection{The West Core\label{sdwest}}

The well defined ring structure of optical and mid--IR emission (Warren--Smith 
\& Scarott \cite{Warren99}; Lefloch \et\ \cite{Lefloch05}) around the West
Core  suggests that its geometry is quite spherical, with a radius of $\sim
0.03$~pc.  The size of the West Core is similar to those from the transient
clumps resolved in L673 (Morata \et\ \cite{Morata05}).

\begin{figure} 
     \begin{center}
    \resizebox{8.0cm}{!}{\includegraphics{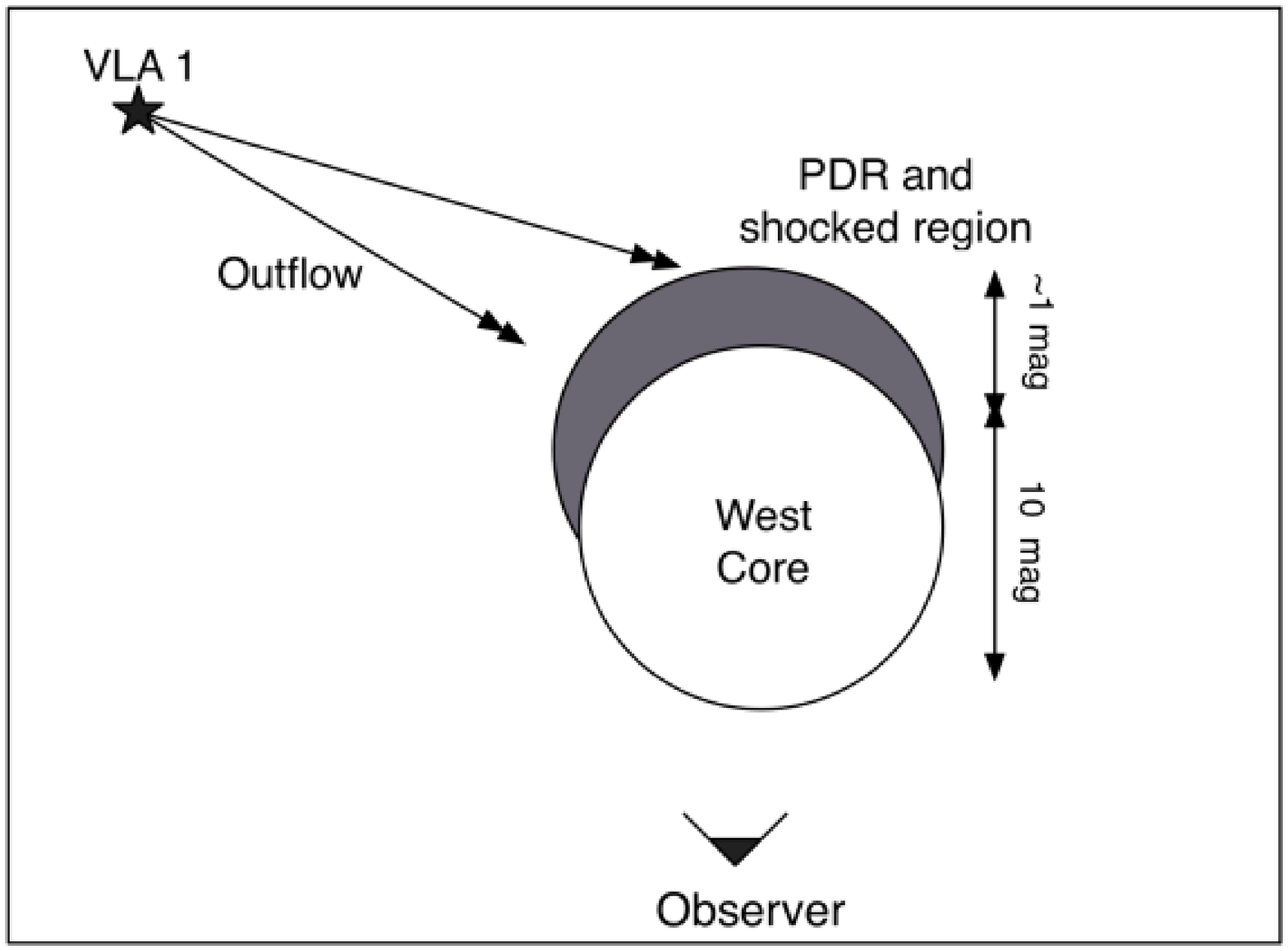}}
    \end{center}
     \caption[]{
Sketch of the West Core, the PDR-like structure, and shocked regions that lie
behind the West Core with respect to the observer point of view.}
     \label{fwest} 
     \end{figure}

The presence of [SII] and H$\alpha$ in the ring indicates that the West Core is
already being shocked by the protostellar winds of the YSO VLA~1. But how can
the wind from VLA~1 form a ring structure around the West Core in the plane of
the sky?  A possible answer is that the shock comes from the `dark face' of the
West Core, as shown in the sketch of Figure~\ref{fwest}. In this case, the cold
and dynamically unperturbed molecular gas and dust is probably masking the
shock where the column densities are high enough.  The gas column density at
the dust peak is  $N($H$_2)\simeq$1\N{22}~\cmd\ (see \S~\ref{sawest}),  which
implies a visual extinction of about 11 magnitudes. This is sufficient to
create an  apparent hole in the [SII] and H$\alpha$ emission at the center of
the West Core.   Following standard extinction laws (\eg\ Mathis
\cite{Mathis90}), the  mid-IR extinction  ranges between $\sim 0.2$ and 0.7
(\ie\ an intensity decrease  between 20\% and 50\%),  which also can account
for the observed  mid--IR ring--like structures.  The \tco, \hco, HCN and CS
spectra from the West Core show two lines, one at a \vlsr\ of $\sim 6.3$~\kms\
and quite narrow (the FWHM of the \tco\ \J{1}{0} line is $\simeq
0.48\pm0.03$~\kms), and a second line, weaker but broader, $\Delta v \simeq
1.5$~\kms, centered at a \vlsr\ of $\sim 8.7$~\kms\ (see Fig.~\ref{fWclump}). 
The narrow line  possibly  traces dynamically unperturbed molecular gas from
the West Core, whereas the broader line may trace a molecular layer of
interaction between the West Core and the shocked material.  Further
observations are required to confirm this hypothesis.

%
%

The ring--like mid--IR emission around the West Core is tracing warm gas
interpreted by Lefloch et al. (\cite{Lefloch05}) as a PDR created by  UV 
radiation field of 20--40 Habing.  Thus, the properties of the West Core,  \ie\
a unperturbed dynamically core surrounded by PDR, suggest that the  West Core
is in a similar situation as the  SO$_2$  Clump.  However, the chemical
composition of the West Core seems different:  the column densities for most
molecules are a factor of $\sim10$ lower in the West Core (Table~\ref{twest})
than in position I4 (Table~\ref{tilumina}), except for HCN, CS and CN which are
of the same order.

The major constraints for the unperturbed cold molecular gas from a chemical
point of view are that i) the line widths are narrow and hence the gas is
probably not shocked; ii) there is an enhancement of HCO$^+$ with respect to a
typical dense core $but$ no other enhancement is observed and in fact CH$_3$OH
and SO are not even detected. The BIMA observations indicate a quiescent, 
possibly cold ($<$ 30K) clump. On the other hand, Lefloch \et\
(\cite{Lefloch05}) detect warm gas interpreted as a  PDR at the edge of this
clump. We have run a PDR model  (with a radiation field ranging from  20 to 40
Habing: Lefloch et al. 2005)  of this region to estimate whether the presence
of a PDR at the edge of the West Core can coexist with the quiescent dense gas
we observe.  We found that as long as the average density of the West Core is
between 5$\times$10$^4$ and 5$\times10^5$ cm$^{-3}$, then it is possible for
the central region of the core to be cold. This range of densities are in
agreement with those derived from the PDR (Lefloch \et\ \cite{Lefloch05}). In
particular we find that for a density of 10$^5$ cm$^{-3}$, the gas temperature
reaches 42 K at 1 mag and 27 K at 3 mags. 

By combining the abundances of the modeled PDR with those from the models in
Paper II we find that the best match is reached by a 2--component structure
made of a fairly old quiescent, photo--processed clump ($\sim$1000 yrs of age)
of thickness less than $\sim 2$~mags plus a PDR component;  such a combination
gives the right abundances of CO, CS, H$_2$CO and CH$_3$OH (coming from the old
dense clump), and of HCO$^+$ (coming mainly from the PDR); we find that the PDR
needs to be at least 1 mags.  

An alternative view is that the mid--IR emission comes not from a conventional
PDR but from a warm interface with a PDR  created in the interaction of the
winds of the YSO VLA1 with the West Core. Nguyen, Hartquist \& Williams
(\cite{Nguyen01}) have explored the physics and chemistry occurring in these
interfaces, and have shown that the thermal pressures in the interfaces should
significantly exceed those within the cloud. Resulting temperatures are
expected to be $\sim 10^3$~K  and dust emissions may therefore mimic those of a
PDR. The results of Nguyen et al. (\cite{Nguyen01}) suggest that, if the
interface occupies a few percent of the core, then some species may achieve
detectable abundances. While CO and CS are not particularly enhanced above cold
cloud abundances, in some conditions HCO$^+$ may be enhanced by up to two
orders of magnitude above its cold cloud abundance. H$_2$CO appears to be
somewhat enhanced; results for methanol are not reported. Thus, it appears
possible that the dynamical interaction of the wind and the clump may lead to a
characteristic chemistry in the interface. Such a model requires more detailed
study, perhaps along the lines of that by Lim, Rawlings \& Williams
(\cite{Lim99}).

\subsection{The High Velocity Region\label{sdxoc}}

The properties of the high velocity \hco\ emission associated with \hhd\ are
similar to those found in the NGC~2071 outflow (Girart \et\ \cite{Girart99}):
there is a good correlation with the shock--excited \h\ (see Fig.~\ref{fxoc})
and the emission has a monotonic enhancement with the flow velocity (Dent \et\
\cite{Dent03}). The high velocity \hco\ emission is also well correlated with
optical knots with small proper motions (\eg\ L, E), with (tangential)
velocities in the 20--80~\kms\ range (Bally \et\ \cite{Bally02}).  Since \hco\
is expected to be strongly enhanced in turbulent mixing layers associated with
low--velocity (40~\kms) shocks (Taylor \& Raga \cite{Taylor95}),  Dent \et\
(\cite{Dent03}) suggested that this flow velocity enhancement dependence could
be explained if the \hco\ is formed at high shock velocities and then gradually
mixes with ambient unenhanced material. This is like the wake chemistry
explored by Lim et al. (\cite{Lim99}) and implies that a UV radiation may not
be necessary in the chemical processing of the \hco\  (Dent \et\
\cite{Dent03}). However, in the mixing layer scenario one would expect  other
molecules, such as CS, \form,  SO, to be also enhanced (\eg\ Viti,  Natarayan
\& Williams \cite{Viti02}), but these are undetected in the High  Velocity
Region. 

Analysis from the mid--IR H$_2$ line emission shows that the  0--0 S($n$) lines
have contributions from hot ($\sim 1000$~K) and warm ($\sim 300$~K)  components
(Lefloch \et\ \cite{Lefloch03}). The hot component arises from  compact regions
($\la 2''$), with shock velocities of 20--30~\kms. The warm  component arises
from more extended and slower shocks (10--15~\kms) in a  denser medium, $\sim
10^5$~\cmt.   The fact that the high velocity \hco\ emission is better
correlated with the  S(2)  \h\ line than with higher excitation \h\ lines and
that this line traces  the warm \h\ component suggest that  the high velocity
\hco\ probably  arises from or around the  \h\  warm component.  

As for the other three regions, we have attempted to  qualitatively model the 
high velocity  gas. The chemical anomaly of this gas is that it differs
substantially from a typical shocked region (see Table~\ref{txoc}).  In
addition, we have tried to see whether UV radiation alone can reproduce the
\hco\ overabundance and found that this is not the case. We have investigated
whether it is possible, via a combination of  heating (e.g. induced by the
passage of a low velocity, non-dissociative shock, as the \hco\ and warm \h\
correlation suggests)  and UV chemistry, for a gas to have such a composition
as we find here, i.e. abundant \hco\ without the presence of other shock
tracers such as SiO and CH$_3$OH.

The models used for this region are similar to those employed in Viti \et\
(\cite{Viti04}) where we explore the chemical evolution of low velocity,
chemically rich clumps observed along the main axis of chemically rich
outflows. The chemical model simulates the clump formation (by free--fall
collapse) and its subsequent interaction with the outflow; here we only
consider the scenario where the density structure formed before the advent of
the outflow (see Viti \et\ \cite{Viti04} for a more detailed explanation of the
model). In addition, here we also include the presence of a strong radiation
field.  After a first stage where densities of up to $10^5$ cm$^{-3}$ are
reached, the region is heated  and affected by a very strong UV field.  We
simulate a short period ($\sim 100$~yrs) of high temperature ($\sim 1000$~K),
followed by cooling, down to a temperature of 200~K. We employ a UV field of 
100  Habing: this value  is derived from the consideration that if the edge of
the West Core is affected by a UV field of 20--40 Habing, then since the  High
Velocity Region is closer to the HH shock by  a factor two  the radiation field
impinging on it   should be about four times stronger.  Table~\ref{txoc}
reports the column densities computed for species observed in this shocked gas.
Observations show that the only detected molecule is HCO$^+$ while for the
other species we have upper limits. We tested two models: Model A where we
assumed a core size of $\sim 0.02$ pc that, at a density of $10^5$\cmt,
corresponds to a visual magnitude of $\sim$ 4.8~mags; and Model B, as Model A
but for a smaller clump of 0.01~pc, corresponding to a visual magnitude of
2.8~mags. We ran the chemistry for 100,000~yrs.

For model A we found that the best match with observations is reached at 
40,000~yrs,  where HCO$^+$ peaks with a fractional abundance of
$5\times10^{-9}$; the gas has by then completely cooled down (to 200~K). For
model B we found that HCO$^+$ peaks earlier,  at 1000~yrs;  the HCO$^+$
fractional abundance is $6\times 10^{-9}$.  For both models, we report in
Table~\ref{txoc} the ratios for selected species to be compared with those
observed and shown in the same table.   If we consider that the HH jet is traveling 
at velocities of hundreds of \kms, then model B seems more realistic because 
of the shorter timescales involved.

There is a remarkable correspondence between the computed peak abundance 
of HCO$^+$, and the observed column density: the latter is derived from 
observations to be $4.1\times10^{13}$ cm$^{-2}$ which implies that for a 
0.01~pc clump (Models B) the observed fractional abundance is 
$\sim 8.9\times 10^{-9}$.

In model B, HCO$^+$ reaches $\sim 10^{-9}$ very early,   at about 500 years;
it  then increases but after about 1500~yrs  starts being destroyed (a
consequence of the lower A$_V$). However, during these 1000~yrs,  which include
the cooled phase, HCO$^+$ is  $\sim 5 \times$10$^{-9}$ for about 400~yrs  and
during that period the ratios of  most of the other species to HCO$^+$ are
consistent with the observed ratios (CO/HCO$^+$ is higher than the ratio estimated 
from the observations by Dent \et\ \cite{Dent03}). In Table~\ref{txoc} 
we report theoretical ratios for all the species, (but not for SiO, which is not 
included in  our models), for which either a detection or an upper limit is 
derived from observations. We chose to show the ratios at the time when HCO$^+$ 
peaks in our models. The species that differ most from observations are HCN 
and CN. As extensively  explained in Paper II, there is a fundamental  problem 
with these two species that at the moment we are unable to explain.

In summary, we have found two suitable scenarios that match the observations of
the  High Velocity  Region  except for the CO, CN, and HCN abundances. The
essential feature is of a  warm  chemistry (possibly due to the
passage of a low velocity, non-dissociative shock)  which is subjected to 
stronger than ambient  UV field, a scenario not previously explored.

\subsection{Is the molecular gas ahead of \hhd\ being driven out by the VLA~1
winds?}

The kinematical properties described in \S~\ref{srcine} suggest that the Ahead
Core is being driven by the protostellar winds from VLA~1. If the velocity 
gradient observed is due to the swept--up effect from the winds, then these
winds should have a low collimation since the dense molecular structure traced
by the \hco\ (including the Ahead Core) is quite extended. Low collimated winds
sweeping up  dense molecular gas have being observed around the HH~34 system
(Anglada \et\ \cite{Anglada95}) and XZ~Tauri (Welch \et\ \cite{Welch00}).

We considered a model of a spherically symmetric stellar wind, with constant
velocity $V_w$ and mass--loss rate $\dot{M}_w$, sweeping up the ambient
material as a snow--plow and accumulating it into a shell.  Anglada \et\
(\cite{Anglada95}) obtains analytical solutions in the case of non--negligible
ambient cloud pressure $\rho\Delta V_c^2$, for ambient cloud density power--law
distributions, $\rho\propto r^{-\alpha}$, with $\alpha=2$ and $\alpha=0$
(constant density). $\Delta V_c$ is the mean squared velocity (turbulent plus
thermal) in the ambient cloud. The parameters of the shell used were those
observed for the Ahead Core, $R_{\rm shell} = 5$\N{4} AU, the distance from the
Ahead Core to VLA~1, $V_{\rm shell} = 3.0$~\kms, the maximum velocity relative
to the systemic velocity observed in the Ahead Core, and  $M_{\rm shell} =
3.8$~\mo, the mass derived from the dust (Dent \et\ \cite{Dent03}), and a
mean--squared velocity $\Delta V_c = 0.6$~\kms, the typical observed line
width. Table~\ref{twind} shows the physical parameters obtained for the density
of the ambient cloud, the age of the shell, the momentum rate of the stellar
wind, and the mass--loss rate for a wind velocity of 100~\kms.

%
     \begin{table}
     \caption[]{Physics of the wind driven Ahead Core$^a$}
     \label{twind}
     \[
\begin{tabular}{lll}
     \noalign{\smallskip}
     \hline
     \noalign{\smallskip}   
\multicolumn{1}{l}{Parameter} &
\multicolumn{2}{c}{Value}  \\
\noalign{\smallskip}
\hline
\noalign{\smallskip}   
$\alpha ^b$                               &  0        &  2 \\
$t_{\rm shell}$ (yr)                      & 3.9\N{4}  & 7.9\N{4}  \\
$n$(H$_2$) (\cmt)                         & 1.3\N{3}  & 4.3\N{2}\,$^c$ \\
$\dot P_w$ (\mo\ yr$^{-1}$ \kms) & 2.9\N{-4} & 1.5\N{-4} \\
${\dot M_w}^d$ (\mo\ yr$^{-1}$)    & 2.9\N{-6} & 1.5\N{-6} \\
     \noalign{\smallskip}
     \hline
     \end{tabular}
     \]
     \begin{list}{}{}
\item[$^a$] Obtained for shell parameters
$R_{\rm shell} = 5$\N{4} AU,
$V_{\rm shell} = 3.0$~\kms, 
$M_{\rm shell} = 3.8$~\mo, and 
$\Delta V_c = 0.6$~\kms.
\item[$^b$] Power index of the the radial density distribution,
$n$(H$_2)\propto r^{- \alpha}$
\item[$^c$] Value for a distance of 5\N{4}~AU.
\item[$^d$] Value obtained assuming a wind velocity of 100~\kms.
     \end{list}
    \end{table}
%

In our case we are only observing the southern part of the region surrounding
the powering source of the HH 1--2 outflow, VLA~1, and the Ahead Core subtends
only a small solid angle $\Omega<4\pi$~sr around VLA~1. However, the results
obtained for a spherically symmetric wind are applicable to the case of a
conical stellar wind.  In general, if the solid angle of the wind cone
$\Omega_w$ is greater than the shell solid angle $\Omega_\mathrm{shell}$, the 
momentum rate, and mass--loss rate obtained in the spherically
symmetric case have to be increased by a factor
$\Omega_w/\Omega_\mathrm{shell}$. $\Omega_w$ is at least 4 times the observed 
$\Omega_\mathrm{shell}$; a factor of 2 comes from the fact that the wind is
bipolar, and an additional factor of 2 comes from the fact that the outflow
axis is very close to the plane of the sky (Solf \& Bohm \cite{Solf91}) and
that the Ahead Core lies in the foreground face of the outflow lobe   (see
\S~\ref{srcine}),  and, for symmetry, the wind has to exist also in the
background face. In conclusion, the  density, momentum rate, and mass--loss
rate obtained in Table~\ref{twind} should be at least  a factor of $\sim4$
higher.

As the outflow axis lies near the plane of the sky, projection effects have to
be accounted for.  The de-projected velocity and distance are 
$V_\mathrm{shell}/\sin i$ and $R_\mathrm{shell}/\cos i$, respectively, where
$i$ is the inclination with respect to the plane of the sky.  However, the
inclination is the biggest uncertainty.   The dynamical age of the HH~1--2
outflow, $\sim1$\N{4} yr ago (Ogura \cite{Ogura95}) and the momentum rate of
the HH 1--2 jet is 8\N{-4} \mo\ yr$^{-1}$ \kms\ (Chernin \& Masson
\cite{Chernin95}). Given the uncertainties, the wind  driven model can roughly
account for these values for inclinations in the range 
30$\arcdeg$--45$\arcdeg$. For such a range $t_{\rm shell}$ does not change much
(a factor 0.6--1) and the outflow momentum rate (for $\alpha=2$) becomes 
$\sim1$--2\N{-3}~\mo\ yr$^{-1}$ \kms.

The total mass of the molecular gas and dust in the HH 1--2 region, derived
from  ammonia observations, is $\sim 52$~\mo\ (Torrelles \et\ \cite{Chema94}).
Taking into account the mass of Ahead Core and of the West Core ($\sim
4.1$~\mo), the fraction of the molecular mass that is being driven out by the
VLA~1 winds in the \hhd\ lobe is $\sim 8$~\%.  Since it is possible that a
similar situation may happen in the northern lobe around HH~1 (Torrelles \et\
\cite{Chema93} found also `quiescent' ammonia clumps ahead of HH~1) the total
mass driven by the powerful VLA~1 is likely 10--15~\% of the total mass in the
HH~1--2 region.     

Finally, it is interesting that while the molecular abundances  may be
explained by chemical models of pre--existing quiescent and transient clumps 
irradiated by a strong UV field (see previous sections and Paper II), the 
kinematics of the regions suggests that the molecular gas is indeed driven out
by  protostellar winds  and the SO line ratios in the SO$_{2}$ clump indicate 
a density increase towards \hhd.   These results seem to be apparently in
contradiction. Although  beyond the scope of this paper and possibly quite
speculative, it is possible that  these clumps have formed due compression by
winds but in a somewhat similar way (though probably faster)  to the formation
of  the transient clumps.  The presence of transient clumps within molecular
clouds has been reported recently both  observationally (Morata \et\
\cite{Morata05}) and  theoretically  (\eg\ Falle \& Hartquist \cite{Falle02};
V\'azquez--Semadeni \et\  \cite{Vazquez05}).

\section{Summary and Conclusions}

We have carried out an extensive observational  study  (from BIMA data) and 
attempted a preliminary theoretical investigation of the molecular gas around
\hhd. The BIMA maps show a very complex  morphological, kinematical and
chemical structure of the molecular gas. For clarity we divided the observed
region in four subregions taking into account the properties of the different
molecular species and the properties observed at other wavelengths:
\begin{itemize}

\item The Ahead Core, located ahead of \hhd, which has a size of $\sim 0.1$~pc
and a mass of 3.8~\mo\ (Dent \et\ \cite{Dent03}). The observed molecular
abundances from the Ahead Core differ from the typical values of low--mass
protostellar envelopes. A weak UV field (weak because attenuated  when passing
through the  SO$_{2}$ clump)  originated in \hhd\ can account for the observed
values if different species trace different layers.   The differences between
species within the Ahead Core suggest that it is not homogeneous but probably
composed by small clumps.

\item The SO$_2$ Clump is a molecular component within the Ahead Core that is
more exposed to the UV radiation from \hhd.   The analysis of the observed
molecular lines indicates an increase of density and relative molecular
abundances towards \hhd.  The photoelectric heating produced by the strong UV
field could probably create a temperature enhancement at the edge of the
clump.  A four point chemical analysis of the  SO$_2$ Clump confirms the
conclusion of Paper II that the clump must have substructure within it  if its
chemistry is due to the \hhd\ UV illumination.  In particular, an increase of
density towards \hhd\ may account for the chemical properties, which would
suggest compression from the VLA~1 winds. From the overall properties of the 
SO$_2$ Clump we cannot say whether it is a clump distinct from the Ahead Core,
or, whether it is part of the face of the Ahead Core exposed to \hhd.

\item The West Core is a molecular structure with a radius of $0.03$~pc and a 
mass of $\sim 0.3$~\mo\ surrounded by a ring structure of shocked gas traced by
[SII] and H$\alpha$ (Warren--Smith  \& Scarott \cite{Warren99}) and of a
PDR-like structure traced by mid--IR hot dust and PAH emission (Lefloch \et\
\cite{Lefloch05}).  This ring--like structure is likely not real but a
consequence of the fact that the West Core is in the foreground with respect 
to the shocked and hot component.  The properties of the West Core and the 
SO$_2$ Clump are somewhat similar (\ie\ apparently quiescent gas irradiated  by
a strong UV field), although their chemistry is different.  We find that the
chemistry of the West Core can be best explained as arising from a combination
of an old photo--processed dense clump, responsible for the emission of CO, CS,
\form\ and \metha, and a PDR, from where most \hco\ emission comes. 
Alternatively, the mid--IR emission may not come from a conventional  PDR but
from a warm interface with PDR created in the interaction of the VLA~1 outflow
with the West Core, which may also account for the \hco\ emission.

\item The High Velocity Region, associated with \hhd, is traced by the \hco\
but not by other molecular shock tracers such as SiO, CS and \metha.  The
presence of enhanced \hco\ and the lack of other shock tracers can be accounted
by the interaction of the VLA~1 outflow with a pre--existing dense clump $via$
heating (possibly due to the passage of a low velocity, non-dissociative
shock)  and by the presence of strong UV field (coming from the nearby high
excitation \hhd\ knots), although our models predict excessive abundances of
HCN and CN.  

\end{itemize}

The overall main conclusion of this work confirms the findings of Paper I and
II,  by demonstrating that in addition to  the strong photochemical effects
caused by  penetration of the UV photons from HH~2 into molecular cloud, a
range of complex radiative and  dynamical interactions occur. These generate
characteristic chemical signatures for each type of interaction.  Thus, despite
the apparent `quiescent' nature of the molecular cloud ahead of \hhd, the
kinematical properties observed within the field of view suggest that the cloud
is not  `quiescent' but it is possibly being driven out by the powerful winds
from the VLA~1 protostar. If so, this would imply that a significant fraction
(10--15~\%) of the total mass in the HH~1--2 region is being disrupted from the
original molecular dense core around the protostar. Given the large size of the
molecular emission, these winds should be of lower collimation  that the jet
associated and responsible of HH 1 and 2.

\begin{acknowledgements}

We thank J. M. Torrelles and S. Curiel for providing the NH$_3$ and [SII]
images, respectively.  We thank the anonymous referee for the thorough review
of the paper.  JMG was supported by RED--2000 from the Generalitat de
Catalunya.  JMG and RE are partially supported by SEUI grant AYA2002-00205.  SV
acknowledges individual financial support from a PPARC Advanced Fellowship. 
DAW thanks the Leverhulme Trust for the award of an Emeritus Fellowship.

\end{acknowledgements}


\clearpage

%
     \setcounter{figure}{0}
     \begin{figure*}
     \resizebox{\hsize}{!}{\includegraphics{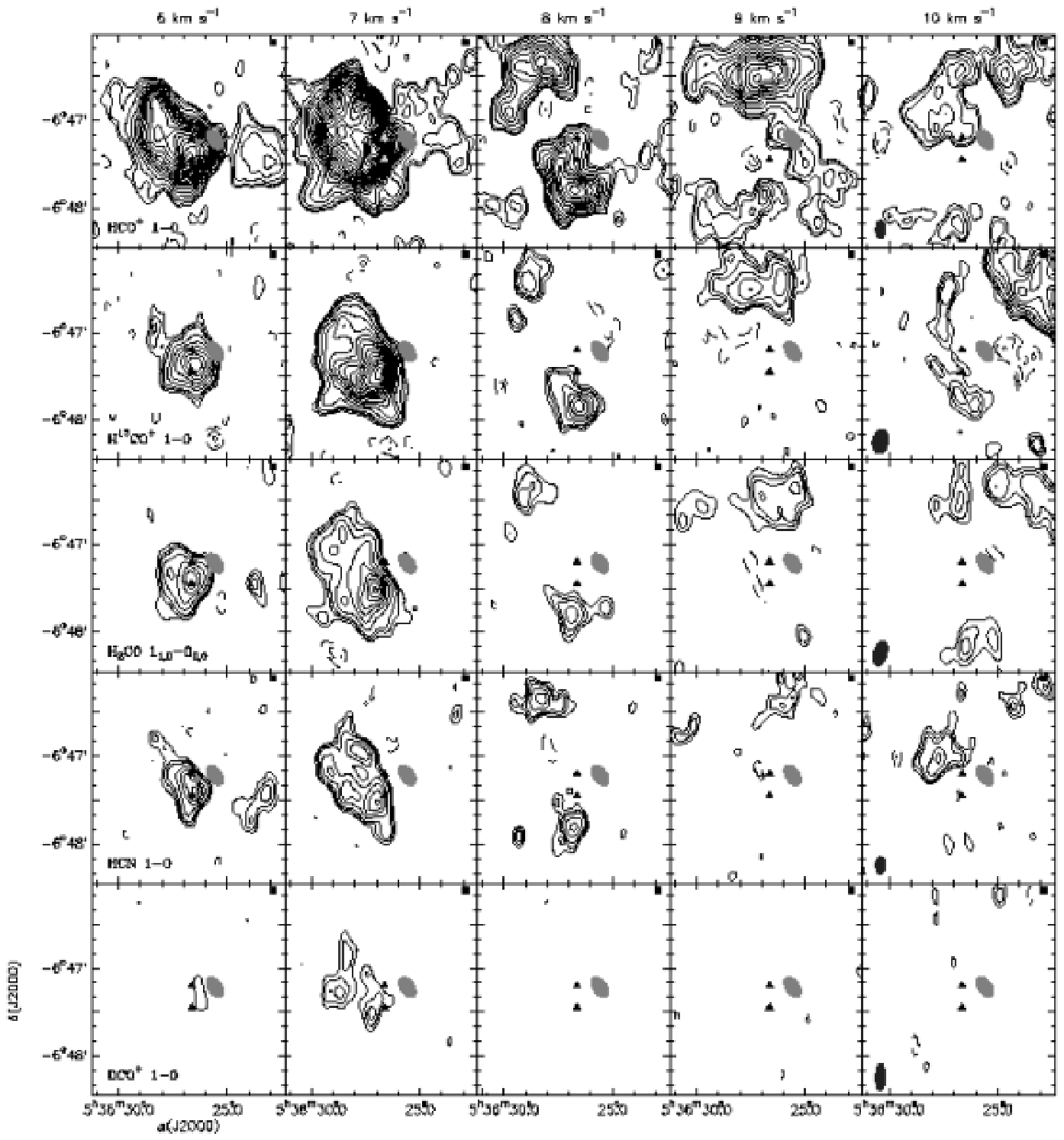}}
     \hfill
     \caption[]{
Channel velocity (for $\Delta v = 1$~\kms) BIMA contour maps of \hco, \htco, 
\form, HCN and DCO$^+$ (the maps are not corrected for primary beam
attenuation).   For the \hco\ the contours are -4, -3, 3, 4, 6, 9, 12, 15, 18,
21, 24, 27, 30, \dots 80 times the $rms$ noise of the map, 62~\mjy. Contours
are -4, -3, 3, 4, 5, 7, 9, 11, \dots\ 31 times the $rms$ noise of the maps: 55
(\htco), 120 (\form), 36 (HCN), and 140~\mjy\ (DCO$^+$).  The CN $N$=1--0  
map was obtained by combining the following hyperfine transitions: 
$F$=5/2--3/2, $J$=3/2--1/2 $F$=3/2--1/2 and $J$=1/2--1/2 $F$=3/2--3/2  
(note that Table~\ref{tbima} give the $rms$ of the maps previous to this 
combination procedure).  The synthesized  beam
is shown in the lower left corner of the right panels.   The grey ellipsoid
shows the position of the brightest HH~2 knots (A, B, D, H). The two filled
triangles show the position of HH~2 knots E (northern one) and L (southern
one). The filled square shows the position  of the driving source of HH~2.
}
     \label{fcanalA} 
     \end{figure*}
%
     \begin{figure*}
     \resizebox{\hsize}{!}{\includegraphics{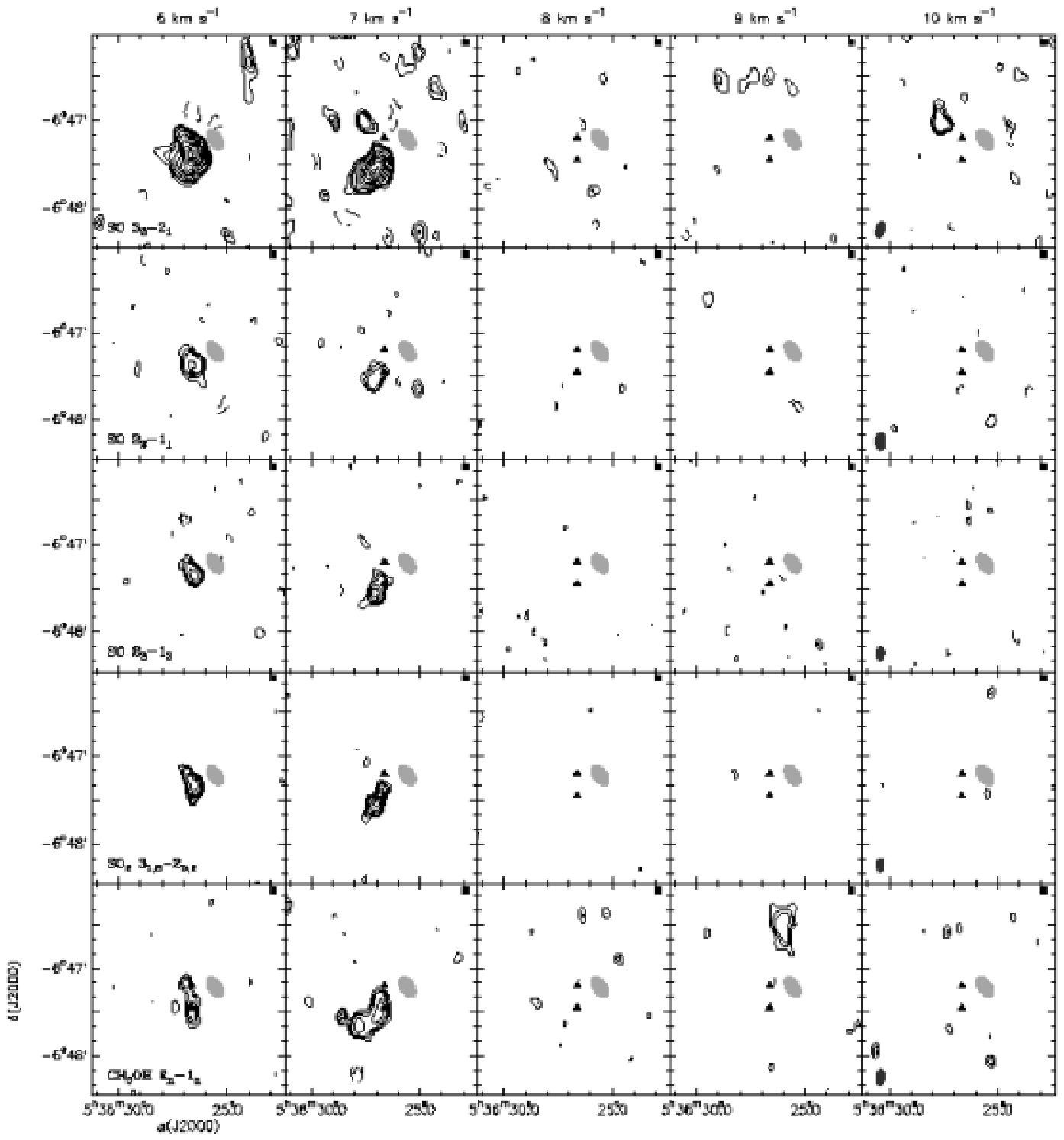}}
     \hfill
     \caption[]{          
Channel velocity BIMA contour maps of three SO lines, SO$_2$ and \metha\  (the
maps are not corrected for primary beam attenuation).  Contours are -4, -3, 3,
4, 5, 7, 9, 11, \dots\ 25 times the $rms$ noise of the maps: 95 (SO 
\JK{3}{2}{2}{1}), 85 (SO  \JK{2}{2}{1}{1}), 73 (SO  \JK{2}{3}{1}{2}),  33
(SO$_2$), and 50~\mjy\ (\metha).  The \metha\  map was obtained by combining 
the \JK{2}{0}{1}{0}\ap\ and \JK{2}{-1}{1}{-1} E transitions (note that 
Table~\ref{tbima} give the $rms$ of the maps previous to this combination 
procedure). 
}
     \label{fcanalB} 
     \end{figure*}
%
     \begin{figure*}
     \resizebox{\hsize}{!}{\includegraphics{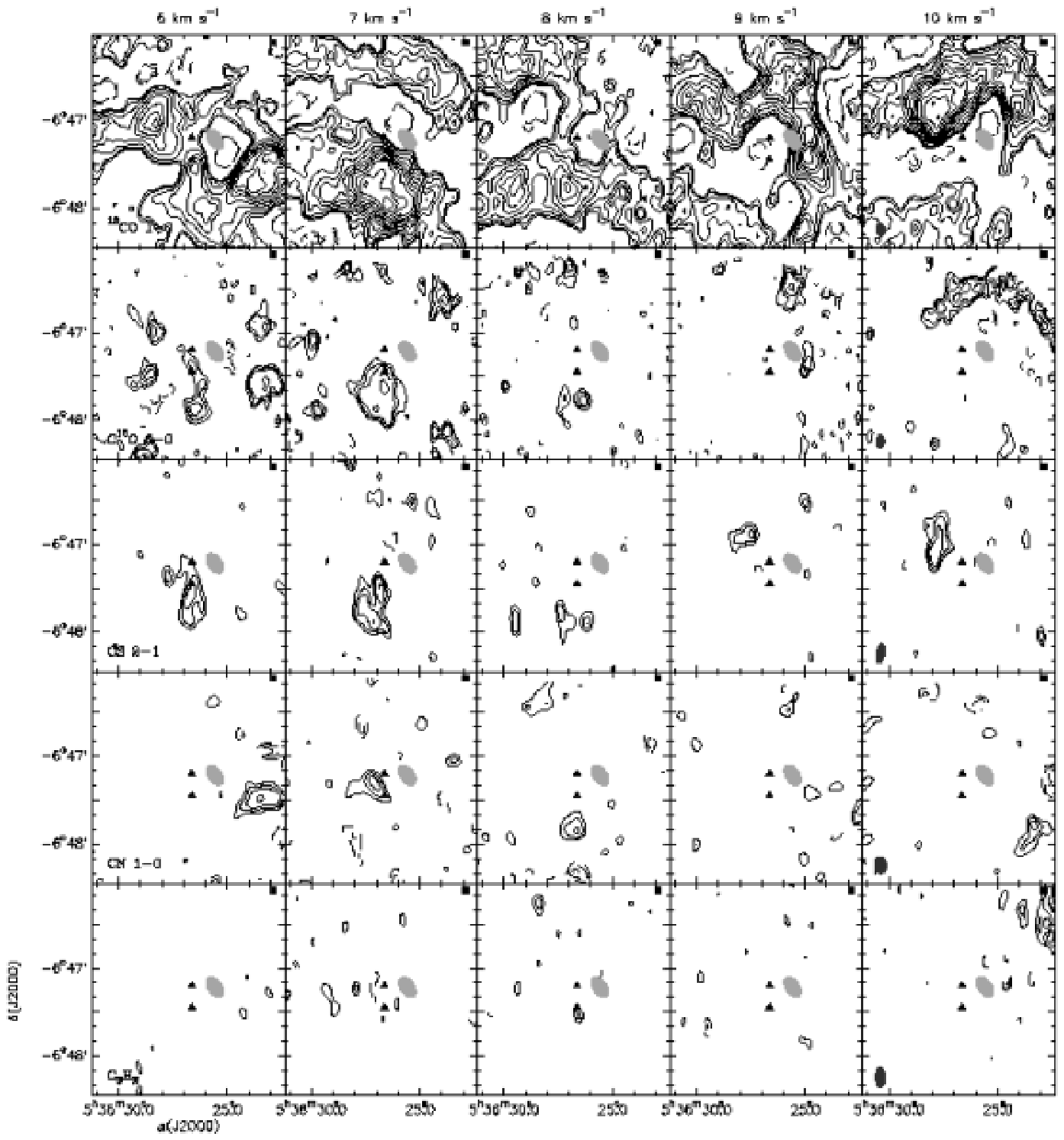}}
     \hfill
     \caption[]{          
Channel velocity BIMA contour maps of \tco, \cdo, CS, CN and \cthd\  (the maps
are not corrected for primary beam attenuation, except for \tco).  Contours are
-4, -3, 3, 4, 5, 7, 9, 11, \dots\ 25 times the $rms$ noise of the maps: 85
(\tco), 75 (\cdo), 83 (CS), 50 (CN), and 52~\mjy\ (\cthd).   The HCN $J$=1--0  
map was obtained by combining the three hyperfine transitions: $F$=2--1, $F$=1--1 
and $F$=0--1  (note that Table~\ref{tbima} give the $rms$ of the maps previous 
to this combination procedure).  The \cdo\ 1--0 maps were obtained combing 
109.7--113.0 and 109.9--113.4~GHz (LSB--USB) frequency setups. 
}
     \label{fcanalC} 
     \end{figure*}
%

\end{document}